\documentclass[referee,a4paper,12pt,traditabstract]{jswsc} 

\usepackage{amsmath}
\usepackage{graphicx}
\usepackage{txfonts}
\usepackage{subfigure}
\usepackage{epstopdf}
\usepackage[displaymath,mathlines]{lineno}
\usepackage[authoryear,round]{natbib}
\usepackage[backref]{hyperref}
\usepackage{url}
\usepackage{xspace}

\newcommand{\td}[2]{\frac{d #1}{d #2}}

\newcommand{\pd}[2]{\frac{\partial#1}{\partial#2}}

\renewcommand{\vec}[1]{\mathbf{#1}}

\renewcommand{\textbf}[1]{\textrm{#1}}

\hypersetup{colorlinks=true,citecolor=cyan,urlcolor=cyan,linkcolor=blue}

\usepackage{xcolor}

\begin{document}

%\begin{linenumbers}  

   \title{Towards advanced forecasting of solar energetic particle events with the PARASOL model}

   %\subtitle{}
   
   \titlerunning{Towards advanced forecasting of SEPs with the PARASOL model}

   \authorrunning{}

   \author{Alexandr Afanasiev\inst{1}
          \and
          Nicolas Wijsen\inst{2}
          \and
          Rami Vainio\inst{1}
          }

   \institute{Department of Physics and Astronomy, University of Turku, FI-20014, Finland\\
            \email{\href{mailto:alexandr.afanasiev@utu.fi}{alexandr.afanasiev@utu.fi}}
        \and
            Centre for mathematical Plasma Astrophysics, KU Leuven Campus Kulak, 8500 Kortrijk, Belgium
        }

%%   \date{Received September 15, 1996; accepted March 16, 1997}

  \abstract
 %% context heading (optional). leave {} empty if necessary  
{Gradual solar energetic particle (SEP) events are generally attributed to the particle acceleration in shock waves driven by coronal mass ejections (CMEs). Space-weather effects of such events are important, so there has been continuous effort to develop models able to forecast their various characteristics. Here we present the first version of a new such model with the primary goal to address energetic storm particle (ESP) events. The model, PARASOL, is built upon the PArticle Radiation Asset Directed at Interplanetary Space Exploration (PARADISE) test-particle simulation model of SEP transport, but includes a semi-analytical description of an inner (i.e., near the shock) part of the foreshock region. The semi-analytical foreshock description is constructed using simulations with the SOLar Particle Acceleration in Coronal Shocks (SOLPACS) model, which simulates proton acceleration self-consistently coupled with Alfvén wave generation upstream of the shock, and subsequent fitting of the simulation results with suitable analytical functions. 
PARASOL requires input of solar wind and shock magnetohydrodynamic (MHD) parameters. We evaluate the performance of PARASOL by simulating the 12 July 2012 SEP event, using the EUropean Heliospheric FORecasting Information Asset (EUHFORIA) MHD simulation of the solar wind and CME in this event. The PARASOL simulation has reproduced the observed ESP event ($E\lesssim 5$~MeV) in the close vicinity of the shock within one order of magnitude in intensity.
} 

\keywords{}

\maketitle
%%
%%________________________________________________________________
\section{Introduction} \label{sec:intro}

Solar energetic particle (SEP) events constitute one of the significant aspects of space weather \citep{Baker2004, VainioEtAl2009}. Of particular importance are gradual SEP events, which are associated with shock waves driven by coronal mass ejections (CMEs) \citep[e.g.,][]{Reames99}, as they provide the highest proton intensities and fluences at MeV energies \citep{DesaiGiacalone2016}. Substantial effort has been addressed worldwide to develop forecasting methods of such SEP events, using empirical and physics-based modeling approaches \citep[a recent review by][and references therein]{Whitman23}. The latter approach includes modeling of particle acceleration at shock waves. These simulations are computationally demanding if done self-consistently, i.e., if the effect of the turbulence amplification by accelerated particles is included. This high computational cost currently prevents such models from being directly used in an operational SEP forecasting framework. Consequently, alternative models have been developed, which approximate the particle acceleration processes through semi-analytical treatments.

An example of such a model is the improved Particle Acceleration and Transport in the Heliosphere \citep[iPATH; e.g.,][]{hu17}. This model includes a particle acceleration module that, at each time step in the simulation, relies on steady-state analytical solutions for diffusive shock acceleration \citep[DSA; e.g.,][]{Drury83} at the shock wave to inject SEPs into the simulation. The utilized DSA solution relies on the local shock properties, which are obtained from a magnetohydrodynamic (MHD) simulation of a CME. The injected particles diffuse within an onion-shell model situated behind the CME-driven shock. Those particles that succeed in escaping upstream of the shock are tracked by iPATH's transport module, which solves the focused transport equation \citep[FTE; e.g.,][]{vandenberg20} under the assumption of a uniform Parker solar wind configuration. Over the years, different versions of the iPATH model have been successfully employed to model several SEP events  \citep[see e.g.,][for some recent examples]{li21,zheyi22}.

A semi-analytical model of the foreshock region of an interplanetary shock driven by a CME was developed by \cite{Vainio14}. Its primary objective was to provide a rapid self-consistent characterization of the foreshock region, including particle intensities and mean free paths across a range of particle energies. The model is built upon a steady-state DSA theory given by \cite{Bell78}, which accounts for self-generated Alfv\'en waves upstream of the shock, and the Coronal Shock Acceleration (CSA) simulation model \citep{VainioLaitinen07, BattarbeeVainio13} capable of simulating coupled ion acceleration and Alfv\'en wave generation in a time-dependent manner in a spatially varying solar wind plasma. CSA simulations agree with Bell's theory near the shock upon reaching the steady state \citep{VainioLaitinen07, AfanasievVainio15}. Therefore, the semi-analytical foreshock model employs, as an analytical framework, Bell's theory with some modifications to account for the finite time span of the acceleration process and extent of the foreshock region. The self-consistent CSA simulations were fit to the obtained analytical framework in order to calibrate its parameters. The model provides a description of the foreshock region in terms of the energetic particle intensity distribution and the pitch-angle scattering mean free path as functions of energy and distance from the shock (along a magnetic field line), provided that the energy spectrum cutoff energy, the ambient plasma, and the magnetohydrodynamic (MHD) shock parameters are given.       

In this paper, we introduce a new particle acceleration and transport model called PARASOL, which is based on the SOLar Particle Acceleration in Coronal Shocks (SOLPACS) model simulating proton acceleration at shocks, coupled with Alfvén wave generation \citep{AfanasievVainio15}, and the PArticle Radiation Asset Directed at Interplanetary Space Exploration (PARADISE) model simulating three-dimensional (3-D) particle transport through the  solar wind\citep{wijsen20b}. SOLPACS treats the wave-particle interactions more accurately than CSA by using the full resonance condition of pitch-angle scattering, and has been recently successfully validated against energetic storm particle (ESP) event observations at 1~au \citep{Afanasiev23}. PARADISE tracks the evolution of energetic particle distributions through solar wind configurations generated by the time-dependent 3-D MHD model known as the EUropean Heliospheric FORecasting Information Asset \citep[EUHFORIA;][]{pomoell18}. Moreover, EUHFORIA can be used to simulate the propagation of CMEs through the solar wind. When sufficiently fast, such a CME will drive a shock wave whose properties serve as input to the PARASOL model. 

\begin{figure}
    \centering
    \includegraphics[width=0.7\textwidth]{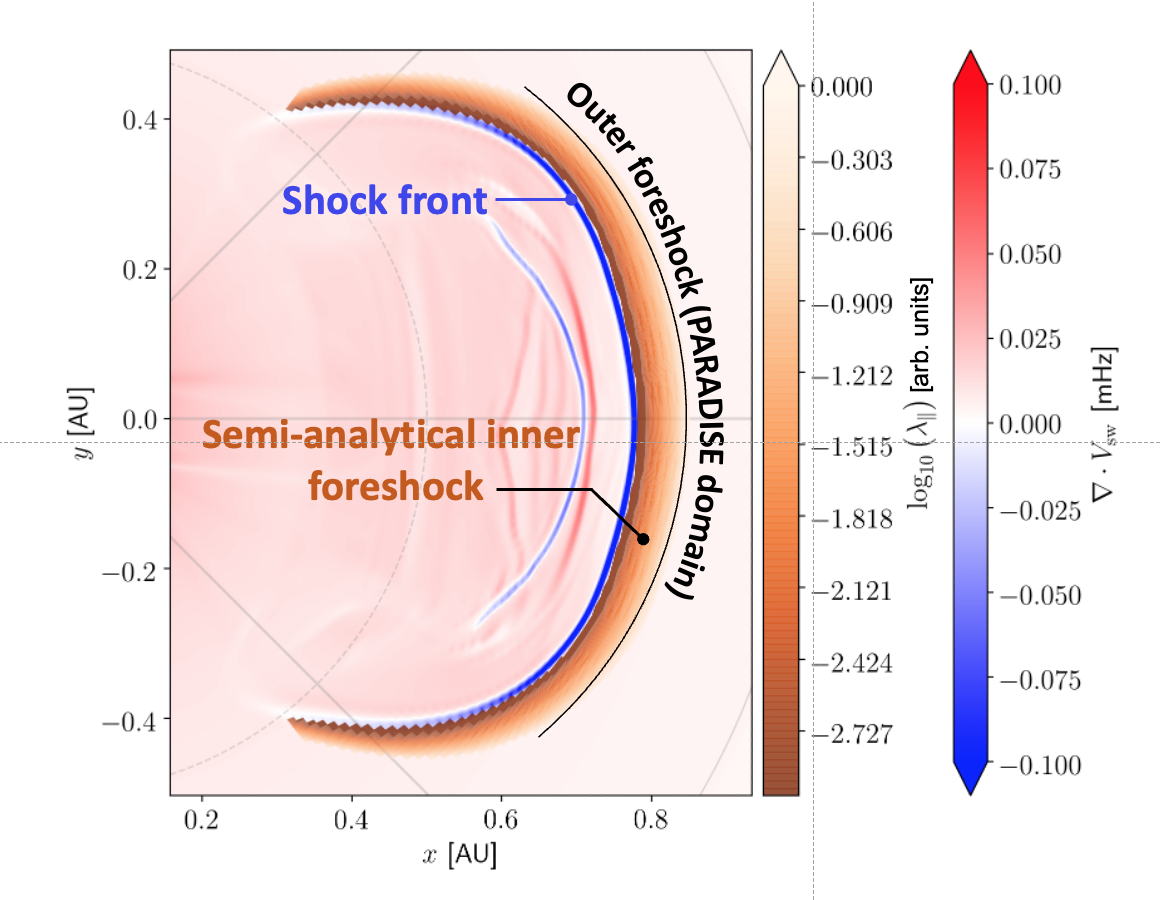}
    \caption{Diagram of the PARASOL model. The thin black line upstream of the shock schematically depicts the interface between the inner foreshock modeled analytically based on self-consistent SOLPACS simulations and the outer foreshock simulated in PARADISE. It should be noted that the actual shock-normal distance from the shock front to the model matching interface depends on the shock magnetic geometry (see Section~\ref{sec:model} for details).}
    \label{fig:parasol_schematic}
\end{figure}

\begin{figure}
    \centering
    \includegraphics[width=1.0\textwidth]{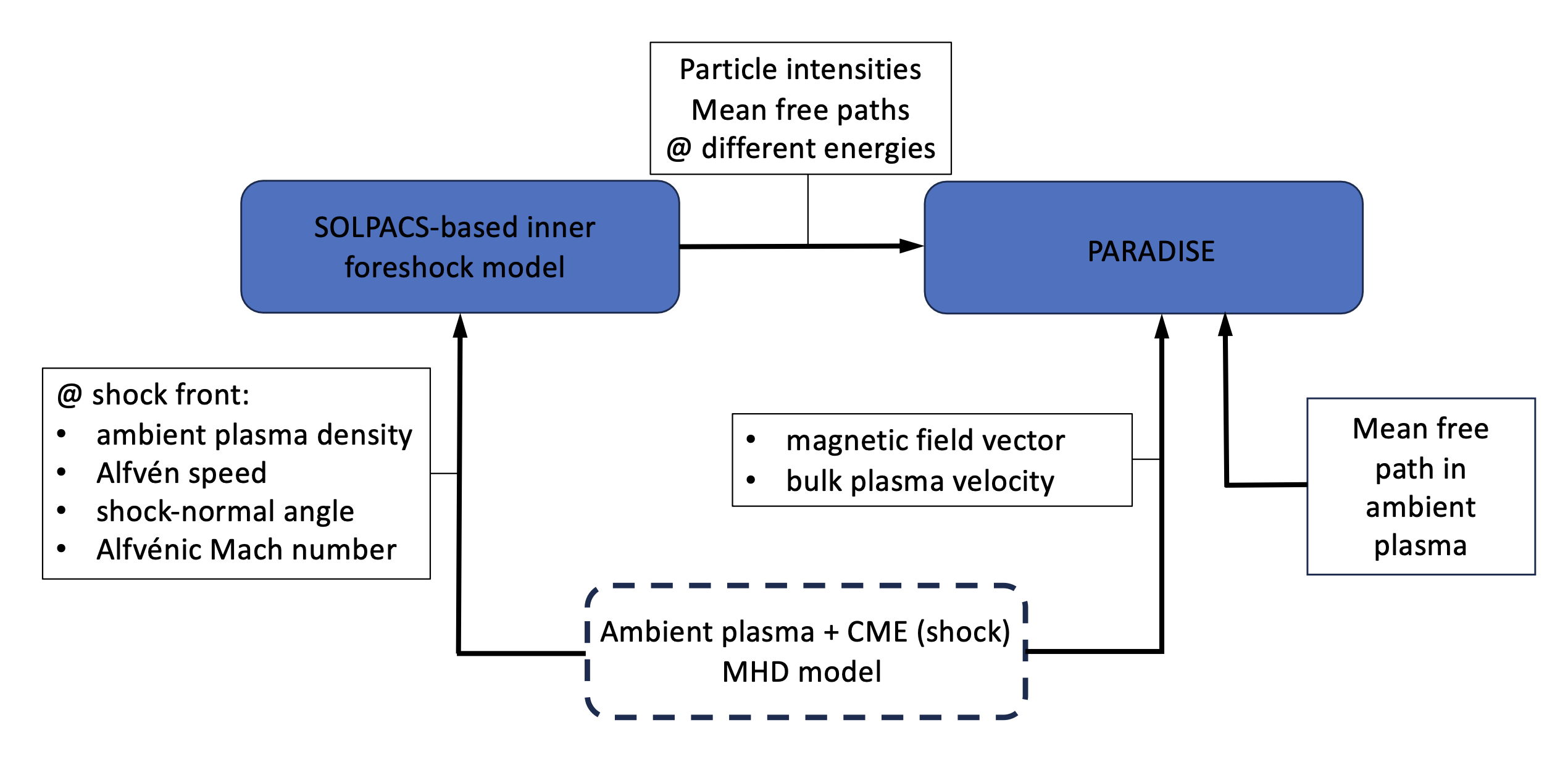}
    \caption{Flow chart of PARASOL indicating the input required by the constituting models.}
    \label{fig:parasol_flowchart}
\end{figure}

PARASOL includes a new semi-analytical foreshock model because directly integrating SOLPACS with PARADISE would result in a computationally heavy simulation model that is hardly usable in the operational space-weather context. Our approach to construct such a model is different from the approach undertaken in \cite{Vainio14}. This is because Bell's theory can not be used in full since SOLPACS simulations give an inverse (compared to Bell's theory and CSA) dependence of the particle mean free path on energy \citep{AfanasievVainio15}, which results from the full resonance condition of pitch-angle scattering utilized in the model. Furthermore, SOLPACS is a local model as it performs simulations under assumption of spatially constant ambient plasma and shock parameters, so no focusing effects are included. As a result, the semi-analytical model developed here describes only the inner part of the foreshock (Fig.~\ref{fig:parasol_schematic}). While the inner foreshock is characterized analytically, the outer foreshock is handled numerically by means of simulations with the PARADISE model, which propagates the accelerated SEP distributions through the EUHFORIA-generated solar wind. 
Figure~\ref{fig:parasol_flowchart} presents a flow chart of PARASOL indicating the input required by the constituting models. It should be noted in this connection that PARASOL can be used not only with EUHFORIA, but with any other model that can provide the necessary bulk plasma and shock parameters.

The paper is structured as follows. In Section~\ref{sec:model}, we outline the derivation of the semi-analytical model and its integration with PARADISE. In Section~\ref{sec:tests}, we present the results of EUHFORIA + PARASOL simulations of an SEP event that was observed in-situ in July 2012.  In Section~\ref{sec:discussion}, we discuss in detail some deviations of the simulation results from observations, as well as further potential developments of the model. Finally, in Section~\ref{sec:Conclusions}, we provide our conclusions.

\section{Model development} \label{sec:model}
\subsection{General consideration} \label{subsec:deriv}

In SEP transport equations that do not address the particle acceleration process at the shock explicitly, the energetic particle emission from the shock is usually described by the source term, which represents the resulting accelerated particle population at the shock \citep[e.g.,][]{La98}. Assuming the shock to be an MHD discontinuity, the source term can be given as 

\begin{equation}\label{eq:S}
S(\mathbf{x}, p, t) = Q(p, t) S_\mathrm{sh}(\mathbf{x},t)
\end{equation}
where $Q(p, t)$ is the particle emission rate as a function of the particle momentum $p$ and time $t$, and $S_\mathrm{sh}(\mathbf{x},t)$ describes the 3-D shock surface at time $t$. For instance, introducing heliocentric spherical coordinates, the latter function can be given as $S_\mathrm{sh}(r, \theta, \phi, t) = \delta(r - r_\mathrm{sh}(\theta, \phi, t) )/(4\pi r^2 \sin{\theta})$, where $r_\mathrm{sh}$ is the radial coordinate of the shock surface at the co-latitude $\theta$ and the longitude $\phi$, and $\delta$ stands for the Dirac's delta function (this particular form of $S_\mathrm{sh}$ assumes that the shock front can intersect a given line drawn from the origin ($\phi = \mathrm{const.}$, $\theta = \mathrm{const.}$) only at one point at a given time).  

The diffusive acceleration of particles at the shock can be described by the 1-D steady-state Parker equation \citep{Drury83}:

\begin{equation}
    u\frac{\partial f}{\partial z}-\frac{p}{3}\frac{\partial u}{\partial z}\frac{\partial f}{\partial p} = \frac{\partial}{\partial z}\left(\kappa\frac{\partial f}{\partial z}\right)+\frac{q}{4\pi p_\mathrm{inj}^2}\delta(z)\delta(p-p_\mathrm{inj}),
    \label{eq:steady_state_parker}
\end{equation}
where $f$ is the isotropic part of the phase-space distribution function of particles, $u$ is the scattering-center speed with respect to the shock and $\kappa = \kappa(z, p)$ is the spatial diffusion coefficient, both measured along the $z$ direction normal to the shock ($z<0$ in the upstream). It is further assumed that seed particles are injected into the acceleration process at a rate $q$ (per unit area of the shock and unit time) at a momentum $p_\mathrm{inj}$. The resulting distribution function at the shock $f_\mathrm{sh} = f(z = 0)$ is then equal to

\begin{equation}
    f_\mathrm{sh}(p) = \frac{3}{4\pi}\frac{q}{\Delta u p_\mathrm{inj}^3}\left(\frac{p}{p_\mathrm{inj}}\right)^{-\sigma},
    \label{eq:distrib_func_shock}
\end{equation}
where $\Delta u = u_1 - u_2$, $\sigma = 3u_1/\Delta u$, and $u_1$ and $u_2$ refer to the shock-normal scattering-center speeds upstream and downstream of the shock, respectively. 

Equation~\eqref{eq:steady_state_parker} can be reformulated to include instead of the seed-particle source term the particle emission term describing the resulting spectrum of accelerated particles:  

\begin{equation}
    u\frac{\partial f}{\partial z}-\frac{\partial}{\partial z}\left(\kappa\frac{\partial f}{\partial z}\right) = Q(p)\delta(z), 
    \label{eq:steady_state_parker_modified}
\end{equation}
where

\begin{equation}
    Q(p)=-\frac{\Delta u}{3}p\frac{\partial f_\mathrm{sh}(p)}{\partial p} 
    \label{eq:emission_rate_parker}
\end{equation}
and $f_\mathrm{sh}(p)$ is given by Eq.~\eqref{eq:distrib_func_shock}.

We employ Eq.~\eqref{eq:emission_rate_parker} to model the particle emission rate at the shock in PARADISE, taking into account that the transport equation solved in PARADISE is formulated in terms of the particle differential intensity $j=p^2f$ \citep[][see also Sec.~\ref{subsec:implem}]{wijsen20b}:   

\begin{equation}
    \begin{split}
        Q^{(P)}(p) &= -\frac{\Delta u}{3}p^{3}\frac{\partial f^{(P)}_\mathrm{sh}(p)}{\partial p} \\
                   &= -U_1 \frac{r_\mathrm{sc}-1}{r} \frac{p^{3}}{3}\frac{\partial}{\partial p} \left(p^{-2} j^{(P)}_\mathrm{sh}\right),
    \end{split}           
    \label{eq:emission_rate_paradise}    
\end{equation}
where $U_1$ is the normal component of the upstream bulk plasma velocity in the de Hoffmann-Teller (HT) frame, $r$ is the conventional gas compression ratio of the shock, $r_\mathrm{sc} = u_1/u_2 = r\left(1 - M^{-1}_\mathrm{A}\right)$ is the scattering-center compression ratio of the shock, $M_\mathrm{A} = U_1/(V_\mathrm{A}\cos{\theta_\mathrm{Bn}})$ is the Alfv\'enic Mach number of the shock as defined in the HT frame, $\theta_\mathrm{Bn}$ is the shock-normal angle, and $V_\mathrm{A}$  is the Alfv\'en speed in the unshocked plasma.  The above expression for the scattering-center compression ratio, $r_\mathrm{sc}$, is obtained under the assumption that upstream of the shock particles are scattered by parallel forward (anti-sunward) propagating Alfv\'en waves (thus, $u_1 = U_1 - V_\mathrm{A}\cos{\theta_\mathrm{Bn}}$) and downstream of the shock particles are scattered by turbulence frozen-in to the bulk plasma \citep[e.g.,][]{Vainio14}. In Eq.~\eqref{eq:emission_rate_paradise} and below, we use the subscript $(P)$ to explicitly acknowledge quantities related to PARADISE (and the subscript $(S)$ to SOLPACS).   

According to Eq.~\eqref{eq:emission_rate_paradise}, the particle emission rate $Q^{(P)}$ is determined by the particle intensity spectrum at the shock, $j^{(P)}_\mathrm{sh}(p)$. To obtain the particle intensity in the foreshock, we need to specify the diffusion coefficient (or the scattering mean free path) in the shock vicinity as a function of momentum (or energy) and distance from the shock. We do this by making use of self-consistent simulations of particle acceleration at shocks performed with the SOLPACS code \citep[][see also Sec.~\ref{subsubsec:solpacs_runs}]{AfanasievVainio15, Afanasiev23}. SOLPACS provides accelerated proton distributions at various energies upstream of the shock, along with their corresponding scattering mean free paths. Due to the self-generated turbulence, the SOLPACS mean free paths in a close vicinity of the shock can be several orders of magnitude smaller than those ($\sim0.1$~au) typically used in PARADISE simulations of particle transport in the interplanetary medium. The direct usage of such small mean free paths would slow down a PARADISE simulation substantially. We deal with this problem in the following way. We require that the particle omnidirectional flux in PARADISE matches the particle omnidirectional flux in SOLPACS only at a certain distance $x_\mathrm{M}$ (along the magnetic field line) upstream of the shock: $j^{(P)}_\mathrm{M}(p) = j^{(S)}_\mathrm{M}(p)$. The PARADISE flux at the shock, $j^{(P)}_\mathrm{sh}(p)$, is then obtained by ``demodulating" the SOLPACS flux at the matching distance $x_\mathrm{M}$, $j_\mathrm{M}^{(S)}(p)$, by using the solution of the steady-state Parker equation:  

\begin{equation}
    j_\mathrm{sh}^{(P)}(p) = j_\mathrm{M}^{(S)}(p)\exp\left\{\int_0^{x_\mathrm{M}} \frac{u_\parallel}{\kappa^{(P)}_\parallel(x,p)} dx\right\},
    \label{eq:PARADISE_shock_spectrum_demodulation}
\end{equation}
where 
$x=-z/\cos\theta_{Bn}$ is the distance along the magnetic field line with respect to the
shock ($x>0$ in the upstream region), $u_\parallel = u_1/\cos{\theta_\mathrm{Bn}}$ and 

\begin{equation}
    \kappa^{(P)}_\parallel(x,p) = \frac{\kappa^{(P)}}{\cos^2{\theta_\mathrm{Bn}}} = \frac{1}{3} v \lambda^{(S)}_{1\,\mathrm{au}}(x,E)
    \label{eq:PARADISE_diffusion_coefficient}
\end{equation}
is the spatial diffusion coefficient along the magnetic field, determined by the parallel mean free path $\lambda^{(S)}_{1\,\mathrm{au}}$ and the particle speed $v$ (to simplify the notation we drop here and below the subscript '$\parallel$' in $\lambda^{(S)}_\parallel$ as in SOLPACS simulation the cross-field diffusion is not considered). For the mean free path in Eq.~\eqref{eq:PARADISE_diffusion_coefficient}, we use an analytical function of distance and energy $E$ that we infer by fitting the mean free paths resulting from SOLPACS simulations for the shock assumed to be located at 1~au. In this way, we can expect PARASOL to be able to predict ESP events measured at 1~au and, at the same time, to be more computationally efficient while simulating particle propagation in the vicinity of the shock (i.e., in the foreshock) when the shock is at distances $<1$~au.

The SOLPACS flux at the matching distance $x_\mathrm{M}$ is obtained via

\begin{equation}
    j_\mathrm{M}^{(S)}(p) = j_\mathrm{sh}^{(S)}(p)\exp\left\{-\int_0^{x_\mathrm{M}} \frac{u_\parallel}{\kappa^{(S)}_\parallel(x,p)} dx\right\},
    \label{eq:SOLPACS_spectrum_at_M}
\end{equation}
with 
\begin{equation}
    \kappa^{(S)}_\parallel(x,p) = \frac{1}{3} v \lambda^{(S)}(x,E),
    \label{eq:SOLPACS_diffusion_coefficient}
\end{equation}
and $j_\mathrm{sh}^{(S)}(p)$ is the particle flux at the shock as modeled by SOLPACS. 

One way to define the matching distance $x_\mathrm{M}$ is to require the residual modulation of the upstream particle flux as modeled in SOLPACS to be small, that is:

\begin{equation}
    M_\mathrm{res} = \int_{x_\mathrm{M}}^{x^0} \frac{u_\parallel}{\kappa^{(S)}_\parallel} dx \ll 1,
\end{equation}
where $x^0$ is the distance along the magnetic field line to the point where the SOLPACS-modeled mean free path becomes equal to the  mean free path $\lambda^0$ used in PARADISE beyond the foreshock (see Eq.~\ref{eq:mfp_upstream}). This allows one to use the larger mean free paths of PARADISE already at $x > x_\mathrm{M}$ in PARASOL, and at the same time to ensure that the escaping particle flux is modeled correctly. As an approximation, we calculate $M_\mathrm{res}$ assuming that $x^0 = x_\mathrm{feb}$, where $x_\mathrm{feb}$ is the distance along the magnetic field line from the shock to the free escape boundary in the upstream, which is equal to the simulation box size in SOLPACS (50~$R_\odot$ in all simulations in this work). Checking the approximated $M_\mathrm{res}$ in our simulations demonstrated that $x_\mathrm{M} > 10~R_\odot$ at energies at which particle intensities and mean free paths in a close vicinity of the shock ($x \lesssim 0.1~R_\odot$) are in a quasi-steady state. Moreover, $x_\mathrm{M}$ is energy-dependent. In this work, we shall treat $x_\mathrm{M}$ as a free parameter of the PARASOL model (and fix it to $10~R_\odot$). 

To summarize this section, in order to construct an analytical model of the foreshock, we need to obtain analytical functions for the particle intensity distribution $j^{(S)}(x, E)$ and mean free path $\lambda^{(S)}(x, E)$ in the upstream region (here and below we use particle energy $E$ instead of momentum $p$). As detailed in the next section, we obtain those by fitting the mean free paths resulting from SOLPACS simulations with analytical functions that represent a modification of the mean free path function of Bell's theory. These fitting functions are then used to evaluate Eqs.~\eqref{eq:PARADISE_shock_spectrum_demodulation} and \eqref{eq:SOLPACS_spectrum_at_M}. This, along with an analytical representation of the simulated intensity spectrum at the shock,  $j^{(S)}_\mathrm{sh}(E)$, allows us to construct the foreshock model based solely on fitting the mean free.

\subsection{Fitting of SOLPACS simulation results} \label{subsec:solpacs_fits}

\subsubsection{SOLPACS simulation runs} \label{subsubsec:solpacs_runs}

In the SOLPACS simulation model \citep{AfanasievVainio15, Afanasiev23}, protons and Alfv\'en waves are traced in a spatially 1-D simulation box (along the magnetic field line) between the shock front and the free-escape boundary placed in the upstream region. The ambient plasma and the magnetic field are assumed homogeneous in the simulation box, and the shock velocity is taken to be constant in a given simulation.
The equations solved by SOLPACS can be given in the following form:
\begin{flalign}
\frac{\partial f}{\partial t}+\left[v\mu+\left(1-M_{\mathrm{A}}\right)V_{\mathrm{A}}\right]\frac{\partial f}{\partial x} & =\frac{\partial}{\partial\mu}\left(D_{\mu\mu}\frac{\partial f}{\partial\mu}\right)\label{eq:solpacs_particle_eq}\\
\frac{\partial I}{\partial t}+\left(1-M_{\mathrm{A}}\right)V_{\mathrm{A}}\frac{\partial I}{\partial x} & =\Gamma\,I\label{eq:solpacs_wave_eq},
\end{flalign}
where Eqs.~\eqref{eq:solpacs_particle_eq} and \eqref{eq:solpacs_wave_eq} describe the evolution of the gyro-averaged distribution function $f(x,v,\mu,t)$ of particles and the evolution of the wave intensity $I(x,k,t)$, correspondingly. Here $t$ is time, $x$ is the spatial coordinate measured along the magnetic field line from the shock toward upstream, $v$ and $\mu$ are the particle speed and the pitch-angle cosine as measured in the rest frame of Alfv\'{e}n waves, and $k$ is the wavenumber. In Eq.~\eqref{eq:solpacs_particle_eq}, the second term on the left-hand side describes particle streaming and convection, and the right-hand side term provides the pitch-angle scattering of particles. In Eq.~\eqref{eq:solpacs_wave_eq}, the second term on the left-hand side describes the wave advection toward the shock, and the term on the right-hand side describes the wave growth. The coefficients $D_{\mu\mu}$ and $\Gamma$ are the quasi-linear pitch-angle diffusion coefficient and the wave growth rate, correspondingly: 
\begin{equation}
D_{\mu\mu}\left(\mu\right)=\frac{\pi}{2}\Omega_{0}\frac{\left|k_{\mathrm{r}}\right|I\left(k_{\mathrm{r}}\right)}{\gamma B^{2}}\left(1-\mu^{2}\right)\label{eq:diff_coef}    
\end{equation}
\begin{equation}
\Gamma\left(k\right)=\frac{\pi}{2}\frac{\Omega_{0}}{nV_{\mathrm{A}}}\int\mathrm{d^{3}}p\,v\left(1-\mu^{2}\right)\left|k\right|\delta\left(k-k_{\mathrm{r}}\right)\frac{\partial f}{\partial\mu}\label{eq:wave_grw},
\end{equation}
where $B$ is the mean magnetic field magnitude, $n$ is the plasma density, $\Omega_0 = eB/m_\mathrm{p}$ is the (non-relativistic) proton cyclotron frequency ($e$ is the electron charge, $m_\mathrm{p}$ is the proton mass), $\gamma$ is the relativistic Lorentz-factor, $k_{\mathrm{r}}$ is the resonant wavenumber given by the quasi-linear resonance condition
\begin{equation}
    k_{\mathrm{r}} = \frac{\Omega_0}{\gamma v \mu},
\end{equation}
and $I\left(k_{\mathrm{r}}\right)$ is the wave intensity at the resonant wavenumber. In Eq.~\eqref{eq:wave_grw}, $\delta\left(k-k_{\mathrm{r}}\right)$ is the Dirac delta-function of its argument, and the integration is performed over the particle momentum space. 

The input parameter list of the SOLPACS model consists of five main parameters: the Alfvén speed $V_\mathrm{A}$ and density $n$ of the ambient plasma, the Alfvénic Mach number of the shock $M_\mathrm{A}$, the shock-normal angle $\theta_\mathrm{Bn}$ and the particle injection efficiency of the shock $\epsilon_\mathrm{inj}$ \citep{Afanasiev23}.

In this study, we used fourteen SOLPACS simulation runs.  The parameter values in each run are given in Table~\ref{tab:sim_params}. The scaling property of the SOLPACS equations, $\alpha^{1/2}\epsilon_\mathrm{inj} = \mathrm{const.}$, where $\alpha$ is the density factor (see Appendix~\ref{appendix scaling property} for details), allows us to narrow down the parameter space. To emphasize this feature, instead of the plasma density $n$ in Table~\ref{tab:sim_params}, we provide the density factor $\alpha$, which is related to the density via $n = \alpha\,n_\mathrm{ref}$, where $n_\mathrm{ref}$ is the reference value of $5\,\mathrm{cm}^{-3}$.    
%
% Table generated by Excel2LaTeX from sheet 'Sheet 1'
\begin{table}[htbp]
  \centering
  \caption{Parameters of the SOLPACS simulation runs. The first column gives the run number, the second the Alfv\'enic Mach number, the third the Alfv\'en speed, the forth the shock-normal angle, the fifth the density factor $\alpha$ (see text for details),  the sixth the injection efficiency, and the seventh the parameter $\sqrt{\alpha}\epsilon_\mathrm{inj}$.  
  }
    \begin{tabular}{rrrrrrr}
    \hline \hline
    Run   & $M_\mathrm{A}$ & $V_\mathrm{A}$~(km/s) & $\theta_\mathrm{Bn}$~($^\circ$) & $\alpha$ & $\epsilon_\mathrm{inj}$ & $\alpha^{1/2}\epsilon_\mathrm{inj}$ \\
    \hline
    1     & 2.5   & 100   & 0     & 100   & $10^{-4}$          & $10^{-3}$ \\
    2     & 5     & 100   & 0     & 100   & $10^{-4}$          & $10^{-3}$ \\
    3     & 5     & 100   & 30    & 100   & $10^{-4}$          & $10^{-3}$ \\
    4     & 5     & 100   & 60    & 100   & $10^{-4}$          & $10^{-3}$ \\
    5     & 10    & 100   & 30    & 100   & $10^{-4}$          & $10^{-3}$ \\
    6     & 10    & 100   & 30    & 100   & $5 \cdot 10^{-5}$  & $5 \cdot 10^{-4}$ \\
    7     & 10    & 100   & 30    & 100   & $5 \cdot 10^{-4}$  & $5 \cdot 10^{-3}$ \\
    8     & 10    & 100   & 30    & 25    & $5 \cdot 10^{-5}$  & $2.5 \cdot 10^{-4}$ \\
    9     & 10    & 100   & 30    & 25    & $5 \cdot 10^{-4}$  & $2.5 \cdot 10^{-3}$ \\
    10    & 10    & 50    & 0     & 25    & $10^{-4}$          & $5 \cdot 10^{-4}$ \\
    11    & 10    & 100   & 30    & 1     & $10^{-3}$          & $1 \cdot 10^{-3}$ \\
    12    & 10    & 20    & 0     & 1     & $10^{-3}$          & $1 \cdot 10^{-3}$ \\
    13    & 10    & 20    & 30    & 1     & $10^{-3}$          & $1 \cdot 10^{-3}$ \\
    14    & 10    & 100   & 30    & 100   & $10^{-3}$          & $1 \cdot 10^{-2}$ \\
    \hline
    \end{tabular}%
  \label{tab:sim_params}%
\end{table}%

\subsubsection{Fitting particle mean free paths and deriving particle intensity distributions}
As outlined above, our approach consists of fitting the simulated mean free paths and applying the steady-state solution of the Parker equation (e.g., Eq.~\eqref{eq:SOLPACS_spectrum_at_M}) to calculate the particle intensity distribution in the upstream region of the shock.

\begin{figure}
    \centering
    \includegraphics[width=0.49\textwidth]{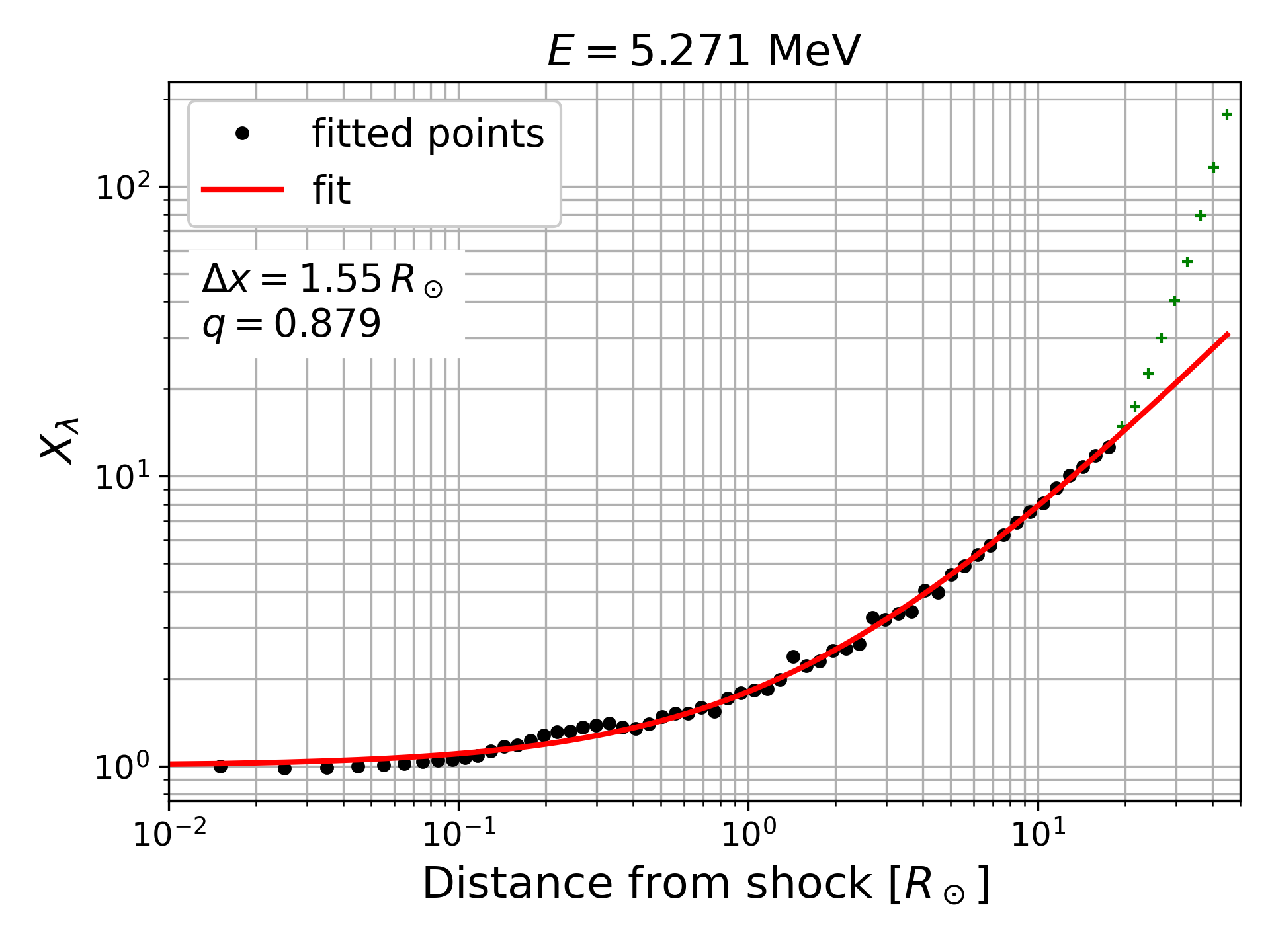}
    \includegraphics[width=0.49\textwidth]{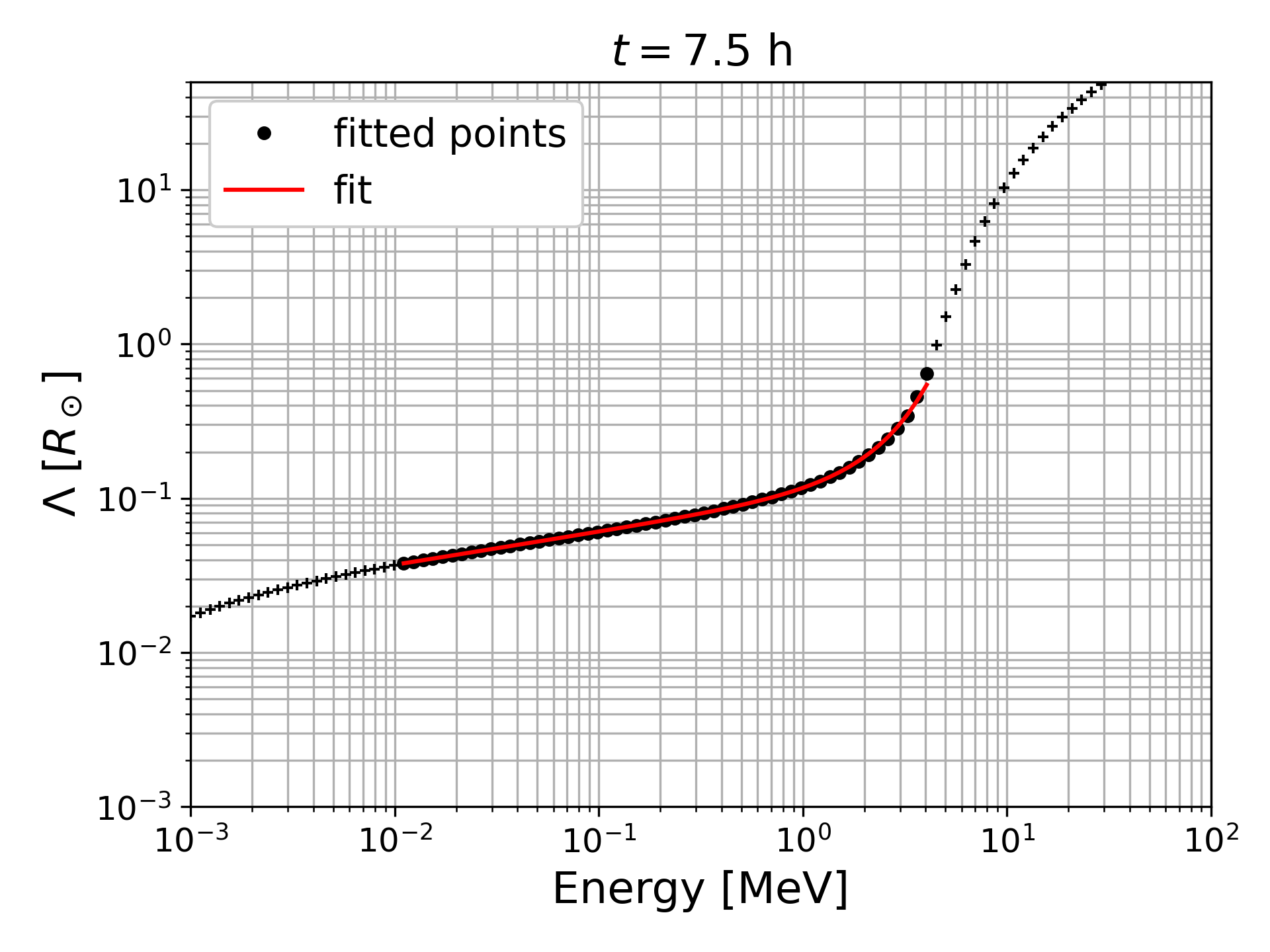}
    \caption{Examples of the fits of the spatial (left panel) and energy (right panel) dependencies of the mean free path in the simulation run 6. The former is obtained at $E = 5.27$~MeV. \textbf{The crosses indicate points that were not considered during the fitting process.} 
    }
    \label{fig:mfp_fits}
\end{figure}

We found that the following function fits well the simulated mean free paths:

\begin{equation}
    \lambda^{(S)}(x,E)=\Lambda(E) X_\lambda(x, E),
    \label{eq:mfp_fitting_function}
\end{equation}
where 
\begin{equation}
\Lambda(E) = K\left(\frac{E}{E_{\mathrm{inj}}}\right)^{\beta_{\lambda}}\exp\left[\left(\frac{E}{E_\mathrm{b}}\right)^{\delta_{\lambda}}\right]
\label{eq:mfp_vs_e}
\end{equation}
gives the mean free path at the shock as a function of energy $E$, and $K$, $\beta_{\lambda}$, $E_\mathrm{b}$ and $\delta_{\lambda}$ are the fitting parameters, and

\begin{equation}
    X_\lambda(x, E) = \left[1+\left(\frac{x}{\Delta x}\right)^{q}\right]^\frac{1}{q}
    \label{eq:mfp_vs_x}
\end{equation}
characterizes the spatial extent of the foreshock, and $\Delta x(E)$ and $q(E)$ are fitting parameters, which are functions of energy. Figure \ref{fig:mfp_fits} shows examples of the fits. 

Equation~\eqref{eq:mfp_vs_e} represents a modification of the energy dependence of the Bell's mean free path at the shock (see Appendix~\ref{appendix: Bell's theory}). We see that the applied fitting function $\Lambda(E)$ describes well the lower-energy part of the mean free path at the shock, which corresponds to the quasi-steady-state power-law part of the particle energy spectrum at the shock. In turn, Eq.~\eqref{eq:mfp_vs_x} presents a modification of the spatial dependence of the Bell's mean free path. The introduced parameter $q<1$ reflects the pitch-angle dependence in the quasi-linear resonance condition of particle scattering. Interestingly, at $x \gg \Delta x$ the mean free path is a linear function of $x$, identical to Bell's theory.    

Using Eqs.~\eqref{eq:mfp_fitting_function} and \eqref{eq:mfp_vs_x}, the particle intensity spectrum at the matching point $x_\mathrm{M}$ in SOLPACS can be calculated as

\begin{equation}
    \begin{split}
        j_\mathrm{M}^{(S)} & = j_\mathrm{sh}^{(S)}\exp\left\{ -\frac{3V_\mathrm{A}(M_\mathrm{A}-1)}{v}\int_{0}^{x_\mathrm{M}}\frac{dx}{\lambda^{(S)}}\right\} \nonumber \\
        & = j_\mathrm{sh}^{(S)}\exp\left\{ -\frac{3V_\mathrm{A}(M_\mathrm{A}-1)}{v\Lambda}\:_{2}F_{1}\left[\frac{1}{q};\frac{1}{q};1+\frac{1}{q};-\left(\frac{x_\mathrm{M}}{\Delta x}\right)^{q}\right]\:x_\mathrm{M}\right\},
    \end{split}
\end{equation}
where $_{2}F_{1}[\cdot]$ is the hypergeometric function \cite[e.g.,][]{AbramowitzStegun_Handbook}. The particle spectrum at the shock to be used in PARADISE
is

\begin{equation}
j_\mathrm{sh}^{(P)} = j_\mathrm{M}^{(S)}\exp\left\{ \frac{3V_\mathrm{A}(M_\mathrm{A} - 1)}{v}\int_{0}^{x_\mathrm{M}}\frac{dx}{\lambda^{(P)}}\right\},
\end{equation}
where
\begin{equation}\label{eq:P_mfp}
\lambda^{(P)}=\lambda^{(S)}_{1\,\mathrm{au}}=\Lambda_0\left[1+\left(\frac{x}{\Delta x_{0}}\right)^{q_{0}}\right]^\frac{1}{q_0},
\end{equation}
the subscript ``0'' indicates that a quantity corresponds to the shock at 1~au. Substituting, we get

\begin{equation}\label{eq:paradise_js}
j_\mathrm{sh}^{(P)} =  j_\mathrm{M}^{(S)}\exp\left\{ \frac{3V_\mathrm{A}(M_\mathrm{A}-1)}{v\Lambda_0}\:_{2}F_{1}\left[\frac{1}{q_{0}};\frac{1}{q_{0}};1+\frac{1}{q_{0}};-\left(\frac{x_{M}}{\Delta x_{0}}\right)^{q_{0}}\right]\:x_\mathrm{M}\right\}. 
\end{equation}

\subsubsection{Particle energy spectrum at the shock}
The particle energy spectrum at the shock simulated in SOLPACS can be reproduced by using the following functional form \citep{AfanasievVainio15}:

\begin{equation}
j_\mathrm{sh}^{(S)}(E)=J\left(\frac{E}{E_{\mathrm{inj}}}\right)^{-\beta}\exp\left[-\left(\frac{E}{E_\mathrm{c}}\right)^{\delta}\right],
\label{eq:part spec at shock}
\end{equation}
where the particle injection energy $E_\mathrm{inj}$, the amplitude factor $J$, and the cutoff (i.e., roll-over) energy $E_\mathrm{c}$ are parameterized as follows: 

\begin{equation}
    E_{\mathrm{inj}}=\frac{1}{2}m_\mathrm{p}M_\mathrm{A}^2V_\mathrm{A}^2\left[\left(3-\frac{2r_\mathrm{mag}}{r}\right)\left(1-\mu_\mathrm{c}\right)+\alpha_\mathrm{inj}\mu_\mathrm{c}\right]^2,
    \label{eq:injection energy}
\end{equation}
where $\mu_\mathrm{c}=\left(1-r_\mathrm{mag}^{-1}\right)^{1/2}$ is the critical value of the pitch-angle cosine, $r_\mathrm{mag}$ is the magnetic compression ratio of the shock, $\alpha_\mathrm{inj} = 4$ is the correction factor,

\begin{equation}
    J=\frac{\beta+1}{\pi\sqrt{8m_\mathrm{p}E_{\mathrm{inj}}}}\epsilon_\mathrm{inj} n_\mathrm{p},
    \label{eq:amplitude factor J}
\end{equation}
\begin{equation}
    E_{\mathrm{c}} = 8.25\cdot10^{-2}\,\left(\frac{E_\mathrm{c, th}}{E_\mathrm{ref}}\right)^{0.86}\,[\mathrm{MeV}],
    %0.857
    \label{eq:cutoff energy fit}
\end{equation}
where

\begin{equation}
    E_\mathrm{c, th}=E_{\mathrm{inj}}\left[1+2\left(\beta_{\lambda}+\frac{1}{2}\right)M_\mathrm{A}^{2}\frac{r_{\mathrm{sc}}(r_{\mathrm{sc}}-1)}{r^2}\frac{V_\mathrm{A}}{v_{\mathrm{inj}}}\frac{V_\mathrm{A}\Delta t}{K}\right]^{1/(\beta_{\lambda}+\frac{1}{2})},
    \label{eq:theory cutoff energy}
\end{equation}
and $E_\mathrm{ref}=1$ MeV.

Equation \eqref{eq:injection energy} is obtained by an ad-hoc modification of the expression for the injection momentum, $p_\mathrm{inj} = 2 m_\mathrm{p} M_\mathrm{A} V_\mathrm{A} (1 - r_\mathrm{mag}r^{-1})$, used by \cite{Vainio14}, whereas Equation~\eqref{eq:amplitude factor J} is derived in the same manner as Eq.~(29) in \cite{Vainio14}.

\begin{figure}
    \centering
    \includegraphics[width=0.49\textwidth]{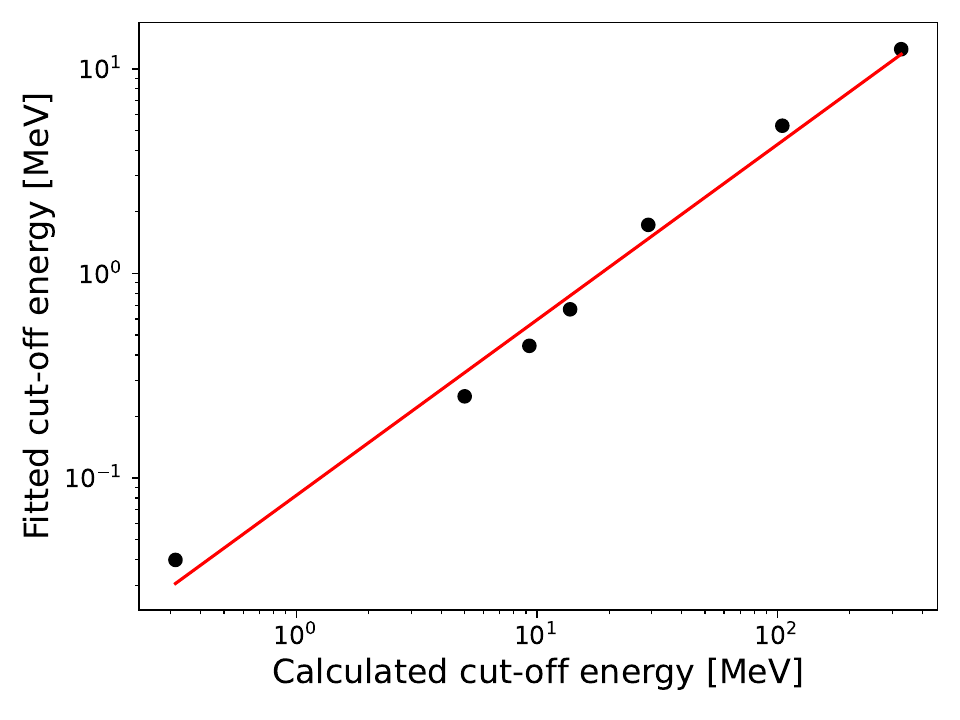}
    \includegraphics[width=0.49\textwidth]{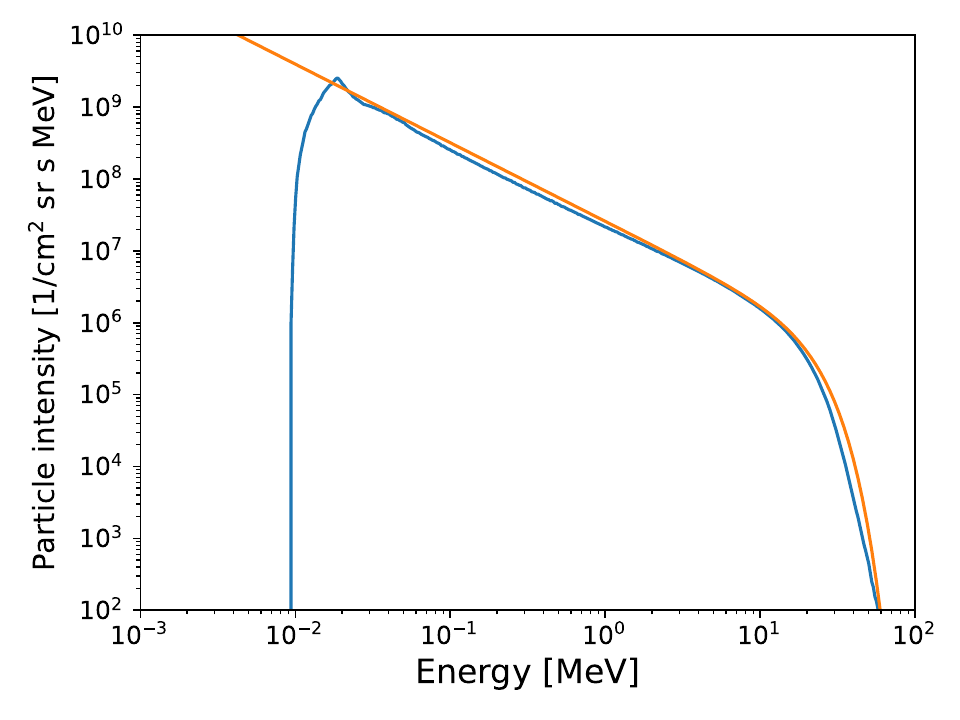}
    \caption{{\it Left panel:} Cut-off energy obtained in Runs 1-7 vs. the cut-off energy calculated based on Eq.~\eqref{eq:theory cutoff energy}. The red line is a power-law fit given by Eq.~\eqref{eq:cutoff energy fit}.
    {\it Right panel:} Example of the simulated particle spectrum at the shock (blue line) and the reproduced spectrum based on the analytical model (orange line). The case presented is run 14.
    }
    \label{fig:pspec_sh_repr}
\end{figure}

Equation~\eqref{eq:cutoff energy fit} is obtained by using an approach similar to \cite{Vainio14}. Namely, we first integrate the analytical expression for the particle acceleration rate \citep{Drury83} over time, assuming a purely power-law dependence for the mean free path at the shock, i.e., $\Lambda(E) = K(E/E_\mathrm{inj})^{\beta_{\lambda}}$ \citep[cf. Eq.~(15) in][]{Vainio14}. This results in Eq.~\eqref{eq:theory cutoff energy}. Then we fit the simulated energy spectra, using Eq.~\eqref{eq:part spec at shock} as a fitting function with $\Tilde{J}$ ($=JE_\mathrm{inj}^\beta$), $\beta$, $E_\mathrm{c}$ and $\delta$ treated as independent fitting parameters. Finally, we compare the theoretical values of the cutoff energy $E_\mathrm{c, th}$ with the fitted ones and fit a power-law function to the resulting dependence (see the left panel of Fig.~\ref{fig:pspec_sh_repr}), which results in Eq.~\eqref{eq:cutoff energy fit}.   

In Eq.~\eqref{eq:theory cutoff energy}, $\Delta t$ is time equivalent to the simulation time in {SOLPACS}. We note, however, that when these equations are implemented in {PARADISE}, it becomes a free parameter that has to be constrained based on test simulations. The right panel of Fig.~\ref{fig:pspec_sh_repr} shows, as an example, a simulated particle energy spectrum at the shock and the reproduced spectrum obtained based on Eqs.~\eqref{eq:part spec at shock}-\eqref{eq:theory cutoff energy}.  

\subsubsection{Obtaining semi-analytical functions for the parameters of the mean free path fitting function}
\begin{figure}
    \centering
    \includegraphics[width=0.8\textwidth]{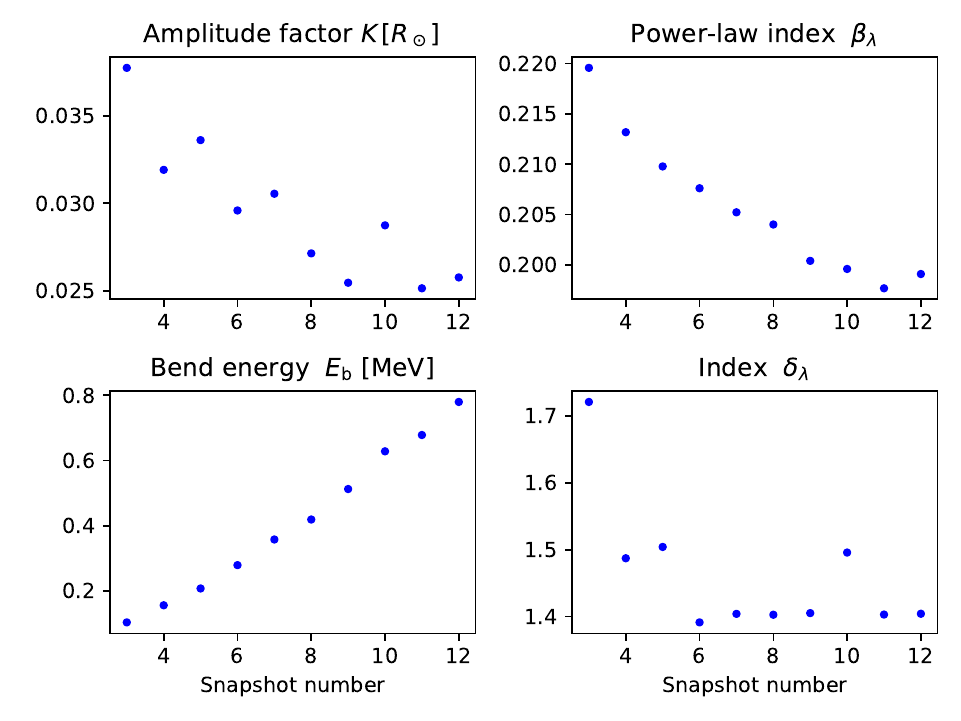}
    \caption{Example of temporal evolution of the parameters of the fitting function $\Lambda(E)$ in a SOLPACS simulation. The plots show each parameter versus a simulation snapshot number. The case presented is run 13.}
    \label{fig:mfp efit params vs time}
\end{figure}
To obtain analytical representations for the fitting parameters $K$, $\beta_{\lambda}$, $E_\mathrm{b}$ and $\delta_{\lambda}$ of the function $\Lambda(E)$ given by Eq.~\eqref{eq:mfp_vs_e}, we studied their evolution with time in several simulation runs. The parameter values were obtained by fitting the energy dependence of the mean free path at the shock at the output times. Figure~\ref{fig:mfp efit params vs time} shows the typical temporal evolution of the parameters in a SOLPACS simulation. One can see that all parameters but one, $\delta_{\lambda}$, exhibit clear systematic changes during a simulation. 

Clearly, the evolution of the amplitude factor $K$ and the power-law index $\beta_{\lambda}$ is due to the contribution of particles being accelerated to progressively higher energies, which then generate waves that resonate with lower-energy particles. Figure~\ref{fig:mfp efit params vs time} shows that, if a simulation runs long enough, the temporal evolution of these parameters slows down. This is expected because there are progressively smaller number of higher-energy particles, modifying the mean free paths of lower-energy particles. Therefore, as the first step, we neglect this temporal evolution at later times and treat these parameters as constants. Specifically, for $\beta_{\lambda}$ we take a value of 0.2.  This value is motivated by our earlier finding that the spectrum of self-generated Alfv\'en waves in SOLPACS simulations approaches asymptotically the $k^{-2}$ form as a function of wavenumber $k$ \citep{AfanasievVainio15}. Taking into account that in the quasi-linear theory the energy dependence of the mean free path is given as $\lambda(E) \propto E^{1-q/2}$, where $q\,~(>0)$ is the power-law index of the wave spectrum, we conclude that the power-law-like part of the energy dependence of the mean free path at the shock should be weak, at least weaker than for the initial $k^{-3/2}$ spectrum, i.e., $\propto E^{1/4}$ \citep[cf. Fig. 7 in ][]{AfanasievVainio15}. 

\begin{figure}
    \centering
    \includegraphics[width=0.6\textwidth]{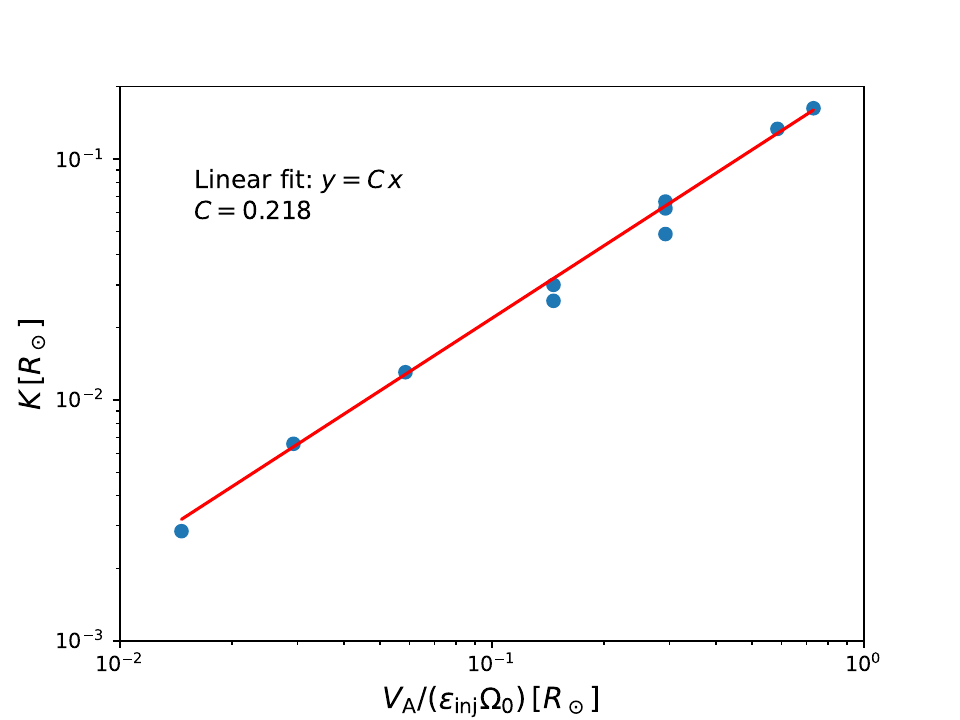}
    \caption{Fit to the amplitude factor $K$ of the mean free path $\Lambda$ at the shock versus $V_\mathrm{A}/(\epsilon_\mathrm{inj}\Omega_\mathrm{0})$.}
    \label{fig:amplitude factor K fit}
\end{figure}

Based on Bell's theory (see Eqs.~\eqref{eq:bell mfp} and \eqref{eq:bell x0} in Appendix~\ref{appendix: Bell's theory}), one would expect the amplitude factor $K$ to scale with $V_\mathrm{A}(\epsilon_\mathrm{inj} \Omega_0)^{-1}$, where $ \Omega_0 $ is the proton cyclotron frequency. We verified this using the full set of simulation runs and found that this relationship holds true (Fig.~\ref{fig:amplitude factor K fit}). Fitting the dependence with a linear function gives:

\begin{equation}
    K = 0.22\,\frac{V_\mathrm{A}}{\epsilon_\mathrm{inj} \Omega_0}.
    \label{eq:result fit for K}
\end{equation}

Regarding the parameter $\delta_{\lambda}$, we fix it at 1.4. It should be noted that $\delta_{\lambda}$ in fact varies at least with $\epsilon_\mathrm{inj}$ \citep[see, e.g., Fig.~7 in][]{AfanasievVainio15}. We will leave this for a detailed investigation in a future study.  

The spatial dependence of the mean free path in the upstream region is fitted using Eq.~\eqref{eq:mfp_vs_x}, in which $\Delta x$ and $q$ are fitting parameters. These parameters are energy-dependent. We found that the following functions provide good fits for the obtained $\Delta x$ and $q$ as functions of energy: 

\begin{equation}
    \Delta x(E) = C_1 \exp{\left(\frac{E}{\Delta E_1}\right)^{\alpha_1}}, 
    \label{eq:Dx_fit_E}
\end{equation}
where $C_1$, $\Delta E_1$, and $\alpha_1$ are fitting parameters, and 

\begin{equation}
    q(E) = C_2 \exp{\left(\frac{E}{\Delta E_2}\right)^{\alpha_2}}, 
    \label{eq:q_fit_E}
\end{equation}
where $C_2$, $\Delta E_2$, and $\alpha_2$ are fitting parameters as well. For the two sets of fitting parameters, we have obtained the following representations: 

\begin{equation}
    C_1 = 0.35 \left(\frac{M_\mathrm{A}}{\epsilon_\mathrm{inj} x_\mathrm{ref}} \frac{V_\mathrm{A}}{\Omega_0}\right)^{0.71}\, [R_\odot],  
    %0.354 0.705
    \label{eq:C1}
\end{equation}
\begin{equation}
    \alpha_1 = 0.83 - 0.02 \frac{E_\mathrm{b}}{E_\mathrm{ref}}, 
    %0.831 0.015
    \label{eq:alpha1}
\end{equation}
\begin{equation}
    \Delta E_1 = 1.64 \left(\frac{E_\mathrm{b}}{E_\mathrm{ref}}\right)^{1.07}\, [\mathrm{MeV}],
    %1.635 1.067
    \label{eq:E1}
\end{equation}
\begin{equation}
    C_2 = 0.66, %0.657
    \label{eq:C2}
\end{equation}
 \begin{equation}
    \alpha_2 = 0.47, %0.467
    \label{eq:alpha2}
\end{equation}
\begin{equation}
    \Delta E_2 = 9.45 \left(\frac{E_\mathrm{b}}{E_\mathrm{ref}}\right)^{0.82} \, [\mathrm{MeV}],
    %9.451 0.822
    \label{eq:E2}
\end{equation}
where $x_\mathrm{ref} = 1\,R_\odot$ and $E_\mathrm{ref} = 1\,\mathrm{MeV}$ are the reference values.

\begin{figure}
    \centering
    \includegraphics[width=\textwidth]{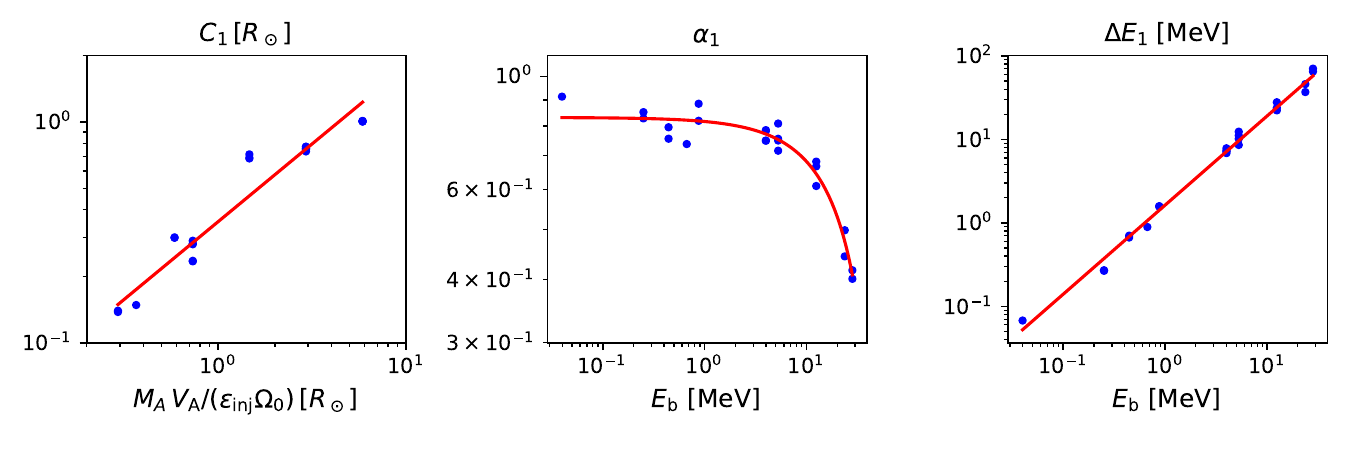}
    \caption{Fits to the parameters of the energy dependence of the foreshock spatial scale $\Delta x$.}
    \label{fig:x0_fit_params_vs_e}
\end{figure}
\begin{figure}
    \centering
    \includegraphics[width=\textwidth]{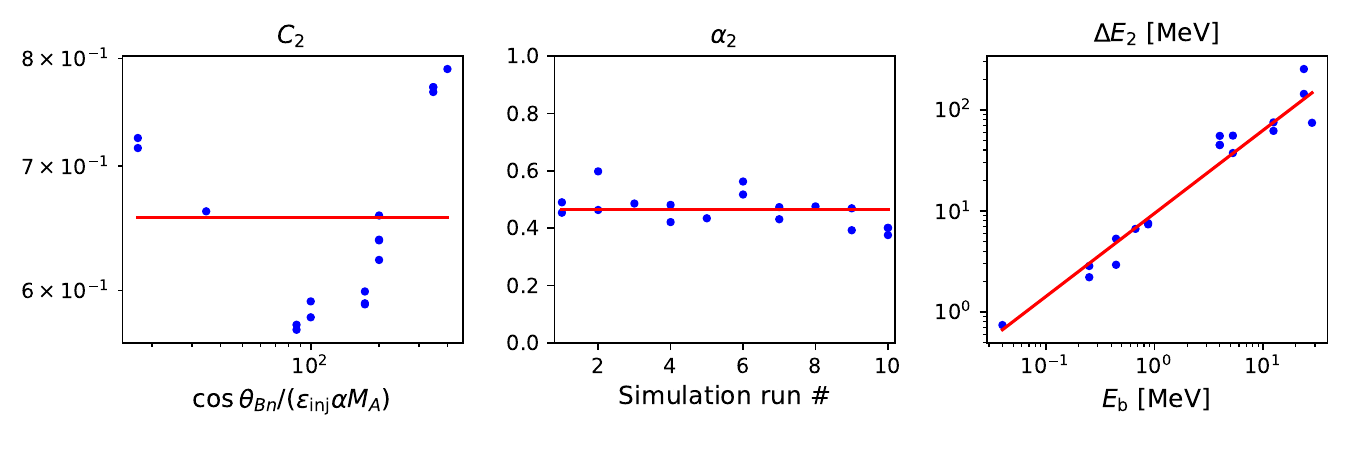}
    \caption{Fits to the parameters of the energy dependence of the foreshock spatial parameter $q$.}
    \label{fig:q_fit_params_vs_e}
\end{figure}

The idea underlying the derivation of Eqs.~\eqref{eq:C1} -- \eqref{eq:E2} for the fitting parameters of $\Delta x(E)$ and $q(E)$ is to make them related to the plasma and shock parameters (and to the model input parameters). 
To obtain Eq.~\eqref{eq:C1} for the fitting parameter $C_1$, which characterizes the foreshock spatial extent $\Delta x$, we used Bell's theory as a starting point since the mean free path at $E \ll E_\mathrm{c}$ is in a quasi-steady state by the end of all the SOLPACS simulations used in this work. In Bell's theory, $C_1 \propto V_\mathrm{A}/(\sigma\epsilon_\mathrm{inj}\Omega_0)$ (see Appendix~\ref{appendix: Bell's theory}). This scaling, however, does not work well with the values of $C_1$ obtained from fitting $\Delta x(E)$ in different SOLPACS simulations. We found that including the Alfve\'nic Mach number into the Bell's scaling organizes the fitted values of $C_1$ into a monotonic dependence that is quite close to a linear dependence (Fig.~\ref{fig:x0_fit_params_vs_e}, left panel). 

It is more difficult to find a suitable combination of plasma and shock parameters for $C_2$ in Eq.~\eqref{eq:q_fit_E} that would organize its values into a monotonic dependence, because $q$ does not exist in Bell's theory. Figure~\ref{fig:q_fit_params_vs_e} (left panel) shows, as an example, the values of $C_2$ versus $\cos{\theta_\mathrm{Bn}}/(\epsilon_\mathrm{inj} \alpha M_\mathrm{A})$. The values, however, vary only within $\sim 20\%$ of their average. Therefore, for $C_2$, we take its average over all the simulations performed in this work.    

The parameters $\Delta E_1$, $\alpha_1$ and $\Delta E_2$ were found to scale with the roll-over energy $E_\mathrm{b}$ of $\Lambda(E)$ (Figs.~\ref{fig:x0_fit_params_vs_e} and \ref{fig:q_fit_params_vs_e}, central and right panels). We used power-law functions to fit $\Delta E_1$ and $\Delta E_2$ versus $E_\mathrm{b}$ and a linear function to fit $\alpha_1$ versus $E_\mathrm{b}$. The parameter $\alpha_2$ was found to be approximately the same in all the simulation runs (Fig.~\ref{fig:q_fit_params_vs_e}, central panel).

\subsection{Implementation in PARADISE} \label{subsec:implem}
The PARADISE model solves the focused transport equation \citep[e.g.,][]{vandenberg20}, that is, 
\begin{align}\label{eq:fte}
\pd{j}{t} &+\nabla\cdot\left(\td{\vec{x}}{t}j\right)
+\pd{}{\mu}\left(\td{\mu}{t}j\right) + \pd{}{p}\left(\td{p}{t}j\right)= 
\pd{}{\mu}\left(D_{\mu\mu}\pd{j}{\mu}\right) + S,
%+ \nabla\cdot\left(\mathbf{D}_\perp \cdot\nabla j\right),
\end{align}
with 
\begin{align}
\td{\vec{x}}{t} =& 
\vec{V}_{\rm sw}+\mu v\hat{\vec{b}}, \label{eq:fte_x}\\
\td{\mu}{t}=&
\frac{1-\mu^2}{2}\left(v \nabla\cdot\hat{\vec{b}} + \mu \nabla\cdot\vec{V}_{\rm sw} - 3 \mu \hat{\vec{b}}\hat{\vec{b}}:\nabla\vec{V}_{\rm sw}- \frac{2}{v}\hat{\vec{b}}\cdot\td{\vec{V}_{\rm sw}}{t} \right),\label{eq:fte_mu}\\
\td{p}{t} =&
 \left(\frac{1-3\mu^2}{2}(\hat{\vec{b}}\hat{\vec{b}}:\nabla\vec{V}_{\rm sw}) - \frac{1-\mu^2}{2}\nabla\cdot\vec{V}_{\rm sw}-\frac{\mu }{v}\hat{\vec{b}}\cdot\td{\vec{V}_{\rm sw}}{t}\right) p, \label{eq:fte_p}
\end{align}
where $j$ denotes the differential particle intensity, $\hat{\vec{b}}$ the unit vector in the magnetic field direction, and $\vec{V}_{\rm{sw}}$ is the solar wind velocity vector. 
These two vector fields are taken from simulations preformed with the time-dependent MHD model EUHFORIA \citep{pomoell18}. 
The source function $S$ is nonzero exclusively at the shock surface, as defined in Eq.~\eqref{eq:S}. The value of $S$ in each point of the shock is obtained by substituting Eqs.~\eqref{eq:paradise_js} and~\eqref{eq:emission_rate_paradise}.
Moreover, for the PARADISE simulations presented in this work, we only incorporate a pitch-angle diffusion process, as described by $D_{\mu\mu}$ in Eq.~\eqref{eq:fte}, while excluding perpendicular diffusion \citep[e.g.,][]{wijsen19a} and guiding center drifts \citep[e.g.,][]{wijsen20a}. The impact of these processes on the PARASOL simulations will be explored in future investigations.

PARADISE uses the results of quasi-linear theory \citep[QLT;][]{jokipii66} to prescribe the following pitch-angle diffusion coefficient \citep[see][]{agueda13,wijsen19a}:

\begin{equation}
    {D_{\mu\mu} = D_0\left(\frac{|\mu|}{1 + |\mu|} + \epsilon\right)\left(1-\mu^2\right)},
\end{equation}
where $\mu$ denotes the cosine of the pitch-angle, $\epsilon = 0.048$ is a parameter bridging the resonance gap at $\mu = 0$ \citep[e.g.][]{klimas71}.
The parameter $D_0$ is determined by prescribing the particles' parallel mean free path $\lambda_\parallel$, which relates to $D_{\mu\mu}$ through \citep{hasselmann70}

\begin{equation}
    \lambda_\parallel = \frac{3 v}{8}\int_{-1}^{1}\frac{\left( 1 - \mu^2\right)^2}{D_{\mu\mu}}d\mu.
\end{equation}
Upstream of the shock wave,  we choose the parallel mean free path 

\begin{align}\label{eq:mfp_upstream}
\lambda_\parallel &= 
\min\left[\lambda^{0}, \lambda^{(P)}\right],
\end{align}
where $\lambda^{(P)}$ is defined in Eq.~\eqref{eq:P_mfp}, and $\lambda^{0} = (0.1~\text{au}) \left({R}/{R_{\rm ref}}\right)^{2-q_0}$. Here, $q_0=5/3$ represents the spectral index of a Kolmogorov turbulence spectrum, and $R$ is the particle rigidity, with $R_{\rm ref}$ set at 43 MV, which corresponds to the rigidity of a 1 MeV proton.

Downstream of the shock, we choose the parallel mean free path as 

\begin{align}\label{eq:mfp_downstream}
\lambda_\parallel &= 
\min\left[\lambda^{0},
\Lambda_0e^{\left(-d/d_0\right)} \right],
\end{align}
where $d$ represents the distance from the shock wave, and we set $d_0 = 1 R_\odot$.  To summarize, Eqs.~\eqref{eq:mfp_upstream} and~\eqref{eq:mfp_downstream} provide the particles with a constant parallel mean free path $\lambda^{0}$ when they are located far from the shock. 
As a particle approaches the shock from the upstream side, it enters a foreshock region where the parallel mean free path, now given by $\lambda^{(P)}$, gradually decreases based on the fittings to the SOLPACS simulations. On the downstream side of the shock, $\lambda_\parallel$ exhibits an exponential increase towards $\lambda^{0}$.

\section{The SEP event of July 12, 2012} \label{sec:tests}
\subsection{Observations}
\begin{figure}
    \centering
    \includegraphics[width=0.99\textwidth]{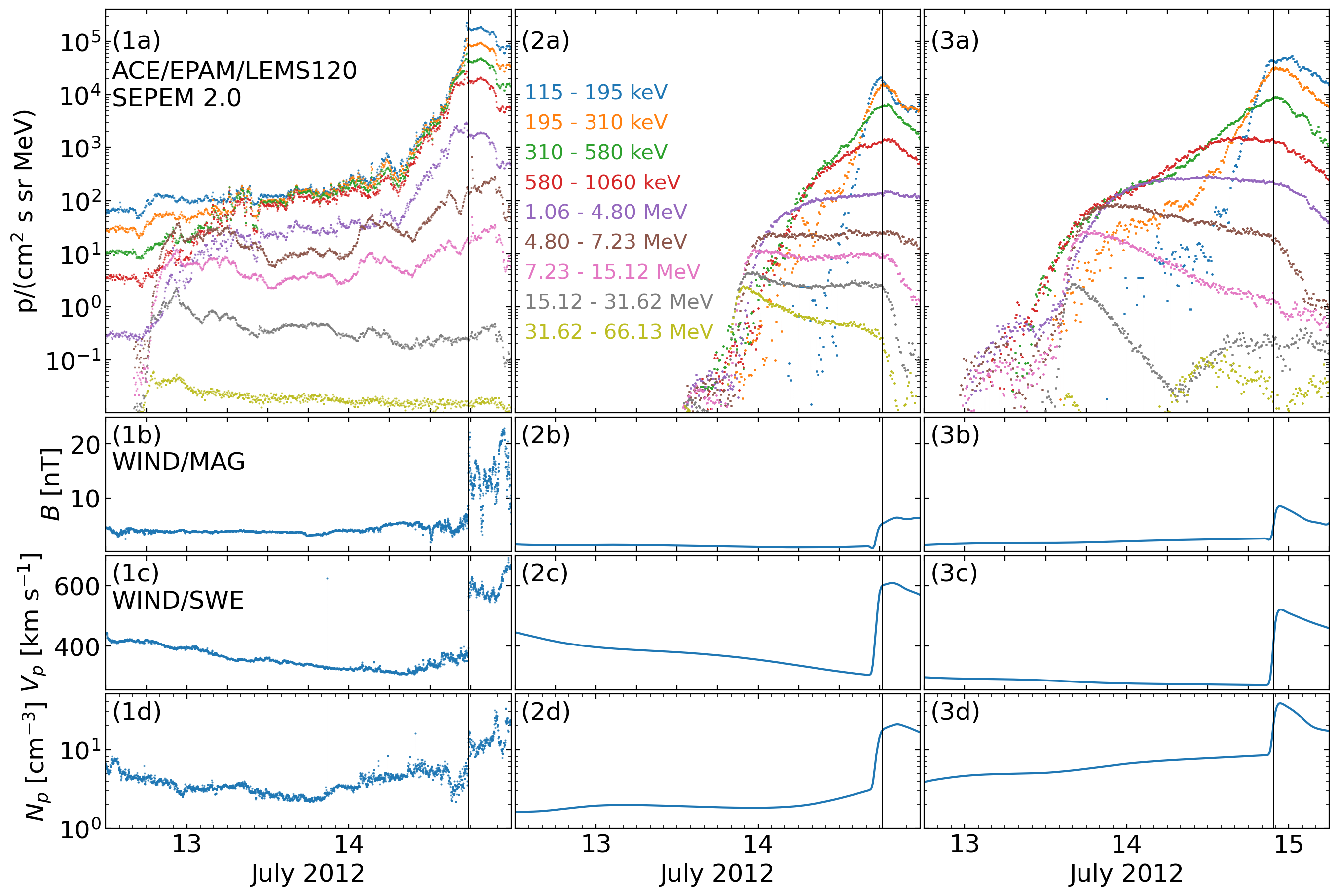}
    \caption{Comparison of observed and simulated proton intensities and solar wind plasma parameters for the 12 July 2012 SEP event. The panels correspond to (a) omnidirectional particle intensities, (b) IMF magnitude, (c) solar wind speed and (d) number density. The left figure displays in-situ observational data obtained from the ACE and WIND spacecraft (refer to the text for further details). In contrast, the central and right figures showcase simulation results for Earth's position and the W25 observer. The vertical lines indicate the arrival of the observed and simulated shock waves at the corresponding observers.}
    \label{fig:obs_and_sim}
\end{figure}

To evaluate the performance of our new model, PARASOL, we conducted simulations of an SEP event observed in situ. Since our solar wind model, EUHFORIA, only starts at 21.5 Rs, this paper will primarily focus on the energetic storm particle (ESP) component of the SEP event. Specifically, we simulated the SEP event associated with a halo CME that occurred on July 12, 2012, and was observed in situ by near-Earth spacecraft. This CME originated from NOAA Active Region 11520, located near the center of the solar disk at the time of the eruption.

In Figure \ref{fig:obs_and_sim}, panel (1a) displays proton intensities in various ion energy channels, as recorded by the Low-Energy Magnetic Spectrometer (LEMS120) of the Electron, Proton, and Alpha Monitor \citep[EPAM;][]{Gold98} aboard the Advanced Composition Explorer (ACE) spacecraft. The panel also includes energy channels from the Solar Energetic Particle Environment Modelling Project (SEPEM) reference data set (RDS) version 2.0 \citep{jiggens18}. The figure illustrates that this SEP event was characterized by an energetic storm particle (ESP) event, which commenced on July 14, 2012, and lasted for approximately a day. The particle intensity peak in the low energy channels coincides with the arrival of the CME-driven shock at Earth, as indicated by the solid line in the figure.

Panel (1b) displays the interplanetary magnetic field (IMF) magnitude measured by the Magnetic Field Investigation (MFI) on Wind \citep{lepping1995}, while panels (1c) and (1d) show, respectively, the solar wind speed and proton number density, as measured by the Solar Wind Experiment (SWE) on Wind \citep{ogilvie1995}. The shock arrival is clearly evident as a sudden jump in all the solar wind variables at 17:26~UT on July 14.

\subsection{EUHFORIA simulation}
The CME responsible for this SEP event was previously simulated using the linear force-free spheromak model within EUHFORIA, with parameters derived from remote-sensing observations \citep{scolini19}. Additionally, the ESP event was recently modeled by \citet{wijsen22} using the PARADISE and EUHFORIA model. In this study, we will utilize the same EUHFORIA simulation as presented in \citet{wijsen22} to describe the solar wind conditions and the propagation of the CME. For an in-depth analysis of the CME and SEP event, as well as the EUHFORIA simulation, we refer to these two papers. To recap the essential parameters, the spheromak CME was introduced into the simulation on July 12, 2012, at 19:24 UT, originating from the inner boundary of EUHFORIA's computational domain, positioned at a heliocentric distance of 0.1~au ($21.5~R_s$). The CME was directed towards Earth and the insertion speed and radius of the spheromak were assumed to be 763 km~s$^{-1}$ and 16.8~$R_\odot$, respectively. A snapshot of the EUHFORIA simulation can be seen in Fig. 3g of \citet{wijsen22}.

\subsection{PARASOL simulation}
\begin{figure}
    \centering
    \includegraphics[width=0.9\textwidth]{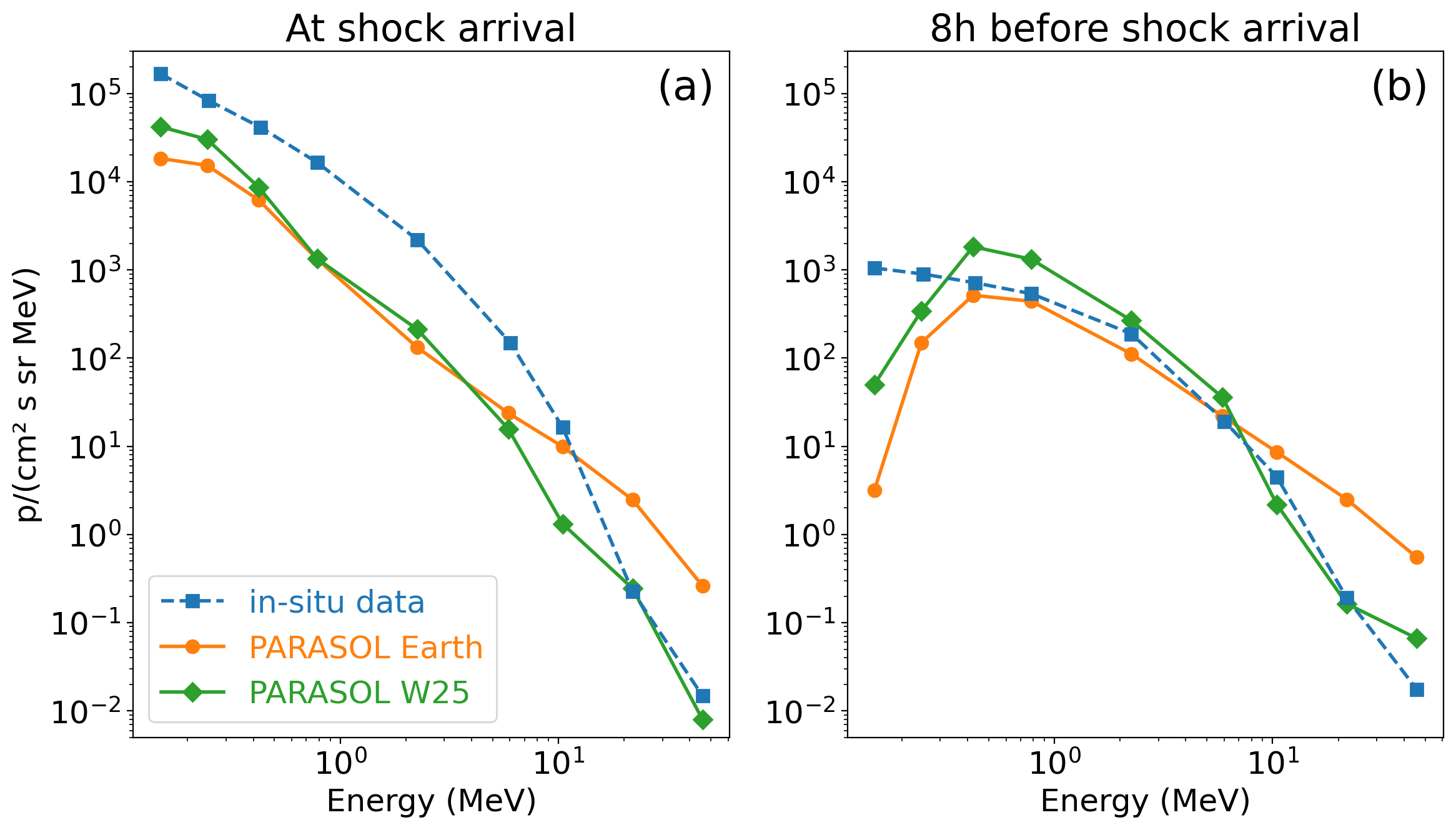}
    \caption{Energy spectra of observed and simulated proton intensities for the 12 July 2012 ESP event. Panel~(a) presents the energy spectrum at the shock arrival, while panel~(b) displays it eight hours before the shock arrival.
    Solid lines represent simulated intensities, with the orange line indicating Earth and the green line representing the W25 observer. Dashed lines correspond to in-situ data from ACE and SEPEM, as also depicted in panel~1a of Figure~\ref{fig:obs_and_sim}.}
    \label{fig:energy_spectrum_sim}
\end{figure}

The PARASOL model comprises three free parameters: the particle injection efficiency $\epsilon_\mathrm{inj}$, the timescale $\Delta t$, and the matching distance $x_m$. As an illustrative example, we present a PARASOL simulation with $\epsilon_\mathrm{inj} = 5 \times 10^{-4}$, $\Delta t = 10$~h, and $x_m = 10 R_s$. A comprehensive parametric investigation into the impact of these three parameters will be detailed in a forthcoming study.

The omnidirectional time-intensity profiles, as modeled by PARASOL, are presented in panels 2a and 3a of Figure \ref{fig:obs_and_sim} for Earth's location and $25^\circ$ west of Earth, referred to as the W25 observer, respectively. Panels 2b - 2d and 3b - 3d display the time profiles of the EUHFORIA solar wind at the locations of these two spacecraft. Observing panel 2a, we note that PARASOL simulates an ESP event. Similar to the observations, the intensity profiles of the lower energy channels peak at the arrival of the modeled CME-driven shock. Moreover, when comparing panels 1a and 2a, it becomes evident that PARASOL slightly underestimates the peak intensity levels in the lowest energy channels and overestimates the intensities in the highest energy channels. In other words, the modeled energy spectrum appears to be harder than the observed one. 
This is illustrated in Figure~\ref{fig:energy_spectrum_sim}a, where the observed and simulated energy spectra at Earth are shown in blue and orange, respectively. The figure highlights that, although PARASOL underestimates intensity at lower energies, the observed and simulated spectra exhibit similar slopes.

Figure \ref{fig:obs_and_sim} also shows that the onset of the SEP event is not accurately reproduced. This discrepancy is a consequence of the spheromak CME not being wide enough to adequately capture the shock wings and the IMF being significantly more radial than the modeled IMF, as elaborated in \citet{wijsen22}. Consequently, in the simulation Earth establishes a magnetic connection with the shock only on July 13, around 22:00 UT. In contrast, the observed onset of the SEP event suggests that Earth likely had a direct magnetic field connection to the shock wave shortly after the CME eruption on July 12, around 17:00 UT.

Due to its more westward position relative to Earth, the W25 observer establishes a magnetic connection to the Earth-directed CME soon after it is inserted into the EUHFORIA simulation at 0.1 au. As a result, the modelled onset of the SEP event for the W25 observer is closer to the observed one. It is still slightly delayed, however, since the computational domain of EUHFORIA, and hence PARASOL, does not include the solar corona. 

Figure~\ref{fig:energy_spectrum_sim}a displays the energy spectrum for the W25 observer at the time of the (simulated) shock arrival. At lower energies, the spectrum closely resembles that of the simulated Earth observer, while at higher energies, the spectrum softens, similar to the observed energy spectrum.  
\begin{figure}
    \centering
    \includegraphics[width=0.6\textwidth]{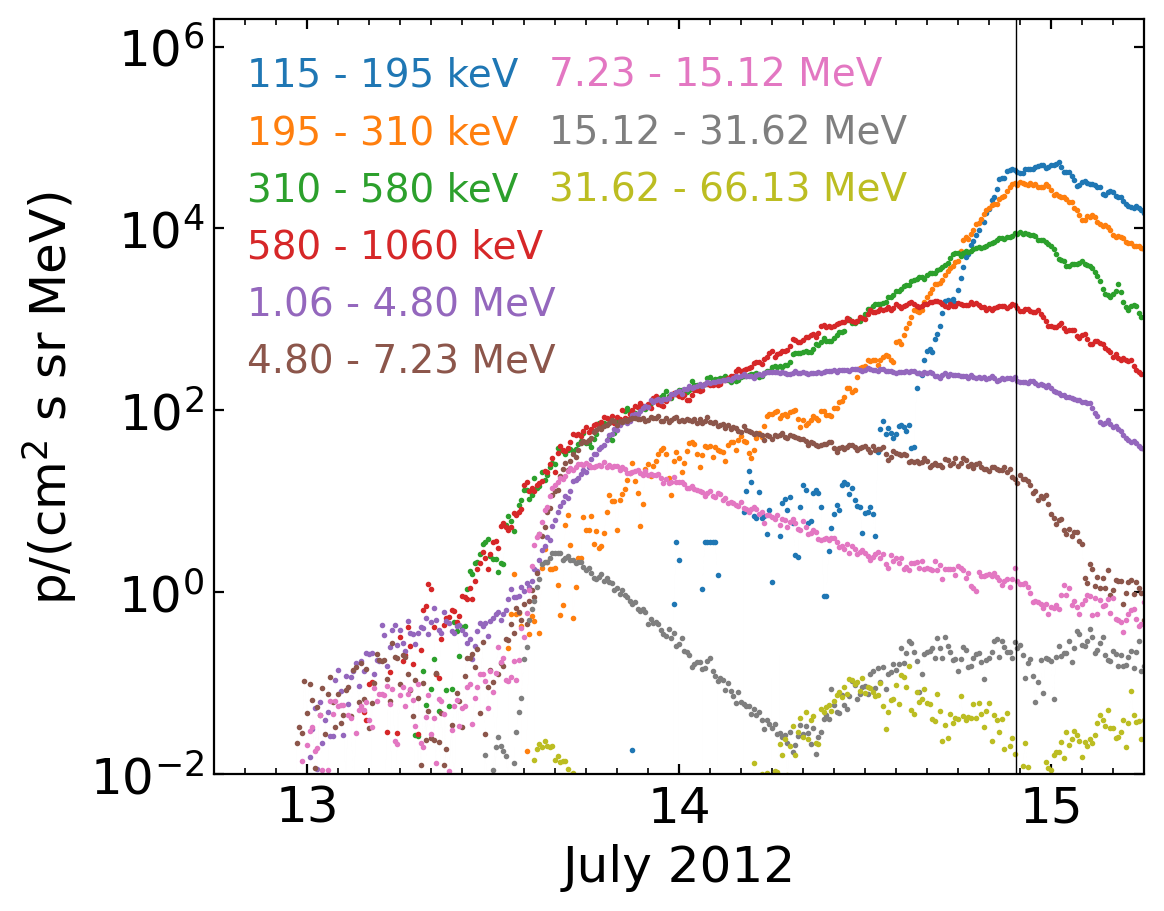}
    \caption{Modeled omnidirectional particle intensities for the W25 observer in a PARASOL simulation with $\epsilon_\mathrm{inj}=10^{-4}$. }
    \label{fig:eps_1e-4}
\end{figure}

As previously mentioned, the intensities in different energy channels are sensitive to the choice of the injection efficiency parameter, $\epsilon_\mathrm{inj}$. This sensitivity is  demonstrated in Figure \ref{fig:eps_1e-4}, which showcases the particle intensities for the W25 observer in a PARASOL simulation with $\epsilon_\mathrm{inj} = 10^{-4}$. A direct comparison between Figure \ref{fig:eps_1e-4} and panel 3a of Figure \ref{fig:obs_and_sim} reveals that a reduced $\epsilon_\mathrm{inj}$ results in lower particle intensities across all energy channels and a less pronounced ESP event. 

To address this sensitivity and determine the optimal value for $\epsilon_\mathrm{inj}$, a future study will involve modeling several different SEP events. This approach will help quantify the most suitable choice for the injection efficiency parameter. In the remainder of this section, we will focus our analysis on the PARASOL simulation with $\epsilon_\mathrm{inj} = 5\times10^{-4}$.

\begin{figure}
    \centering
    \includegraphics[width=0.49\textwidth]{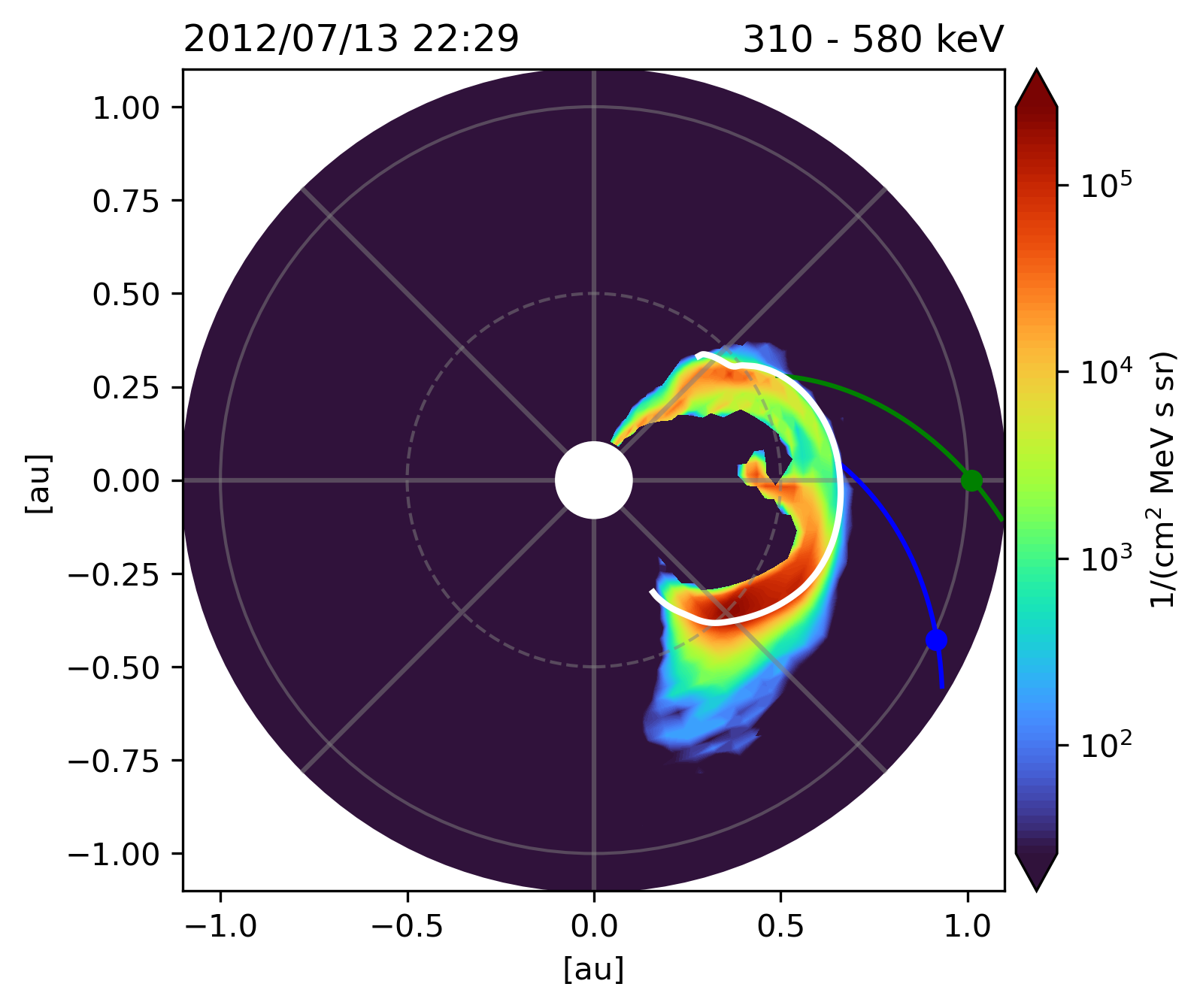}
    \includegraphics[width=0.49\textwidth]{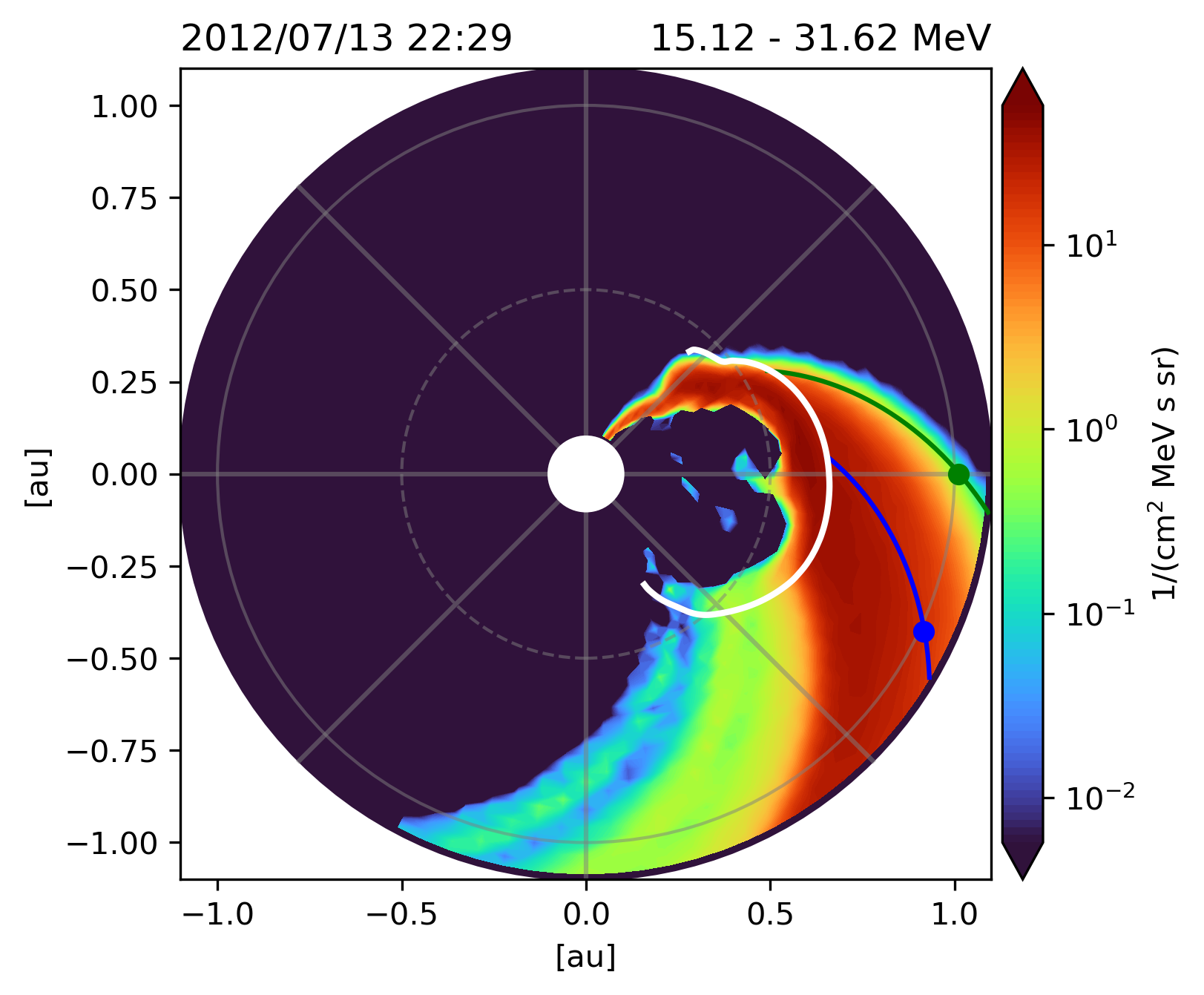}
    \caption{The modeled omnidirectional proton intensities within the plane of constant latitude encompassing Earth, 27h after the insertion of the CME in the simulation domain. The left panel displays proton intensities within the 310–580~keV range, while the right panel presents data for the 15–31~MeV range. The white line denotes the position of the CME-driven shock wave. Earth and the W25 observer are marked by green and blue dots, respectively, and their magnetic connectivity to the shock wave is also illustrated.}
    \label{fig:sim_slice}
\end{figure}

Figure \ref{fig:sim_slice} shows the omnidirectional proton intensities for two different energy channels within the plane of constant latitude encompassing Earth, 27h after the insertion of the CME in the simulation domain.
This figure offers valuable insights into the spatial distribution of proton intensities resulting from the shock's interaction with the surrounding solar wind. In comparing the two panels, we observe that low-energy particles are more tightly confined near the CME than high-energy particles, due to their lower speed (relative to the CME) and their smaller mean free path in the foreshock region, as further discussed in the next section. Furthermore, the right panel reveals that the highest intensities in the high-energy range are concentrated at the CME’s center and west flank, whereas in the low-energy range (left panel), the highest intensities are concentrated along the CME’s east flank.

\begin{figure}
    \centering
    \includegraphics[width=0.49\textwidth]{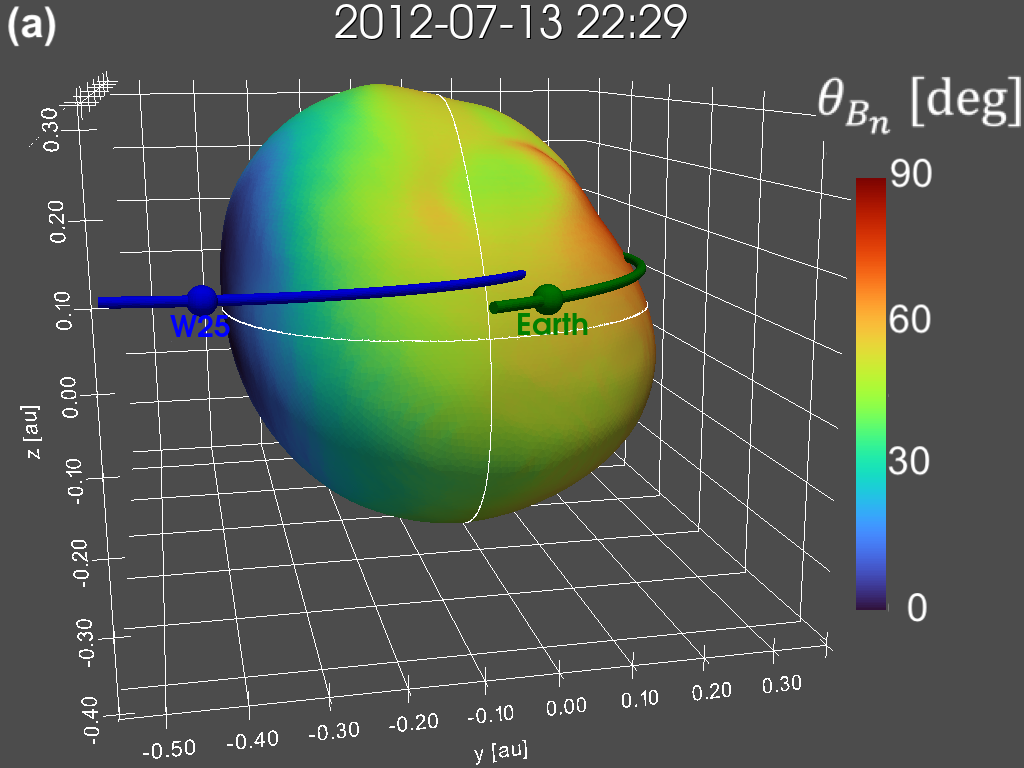}
    \hspace{-2.00mm}
    \includegraphics[width=0.49\textwidth]{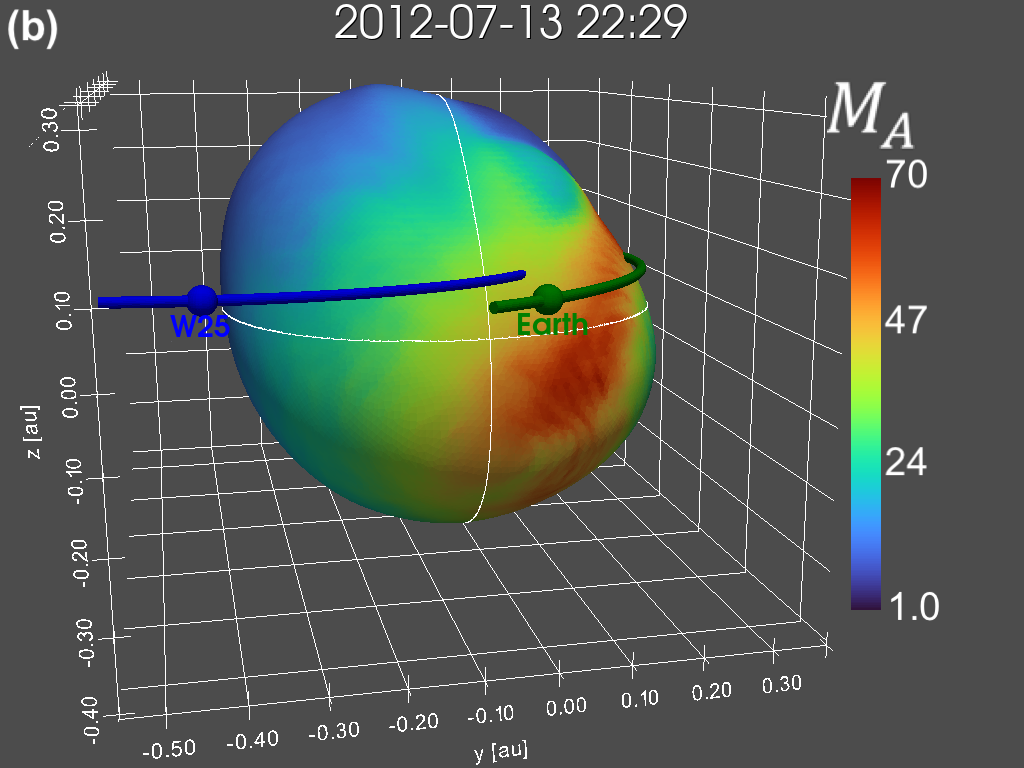}\\
    \vspace{-0.39mm}
    \includegraphics[width=0.49\textwidth]{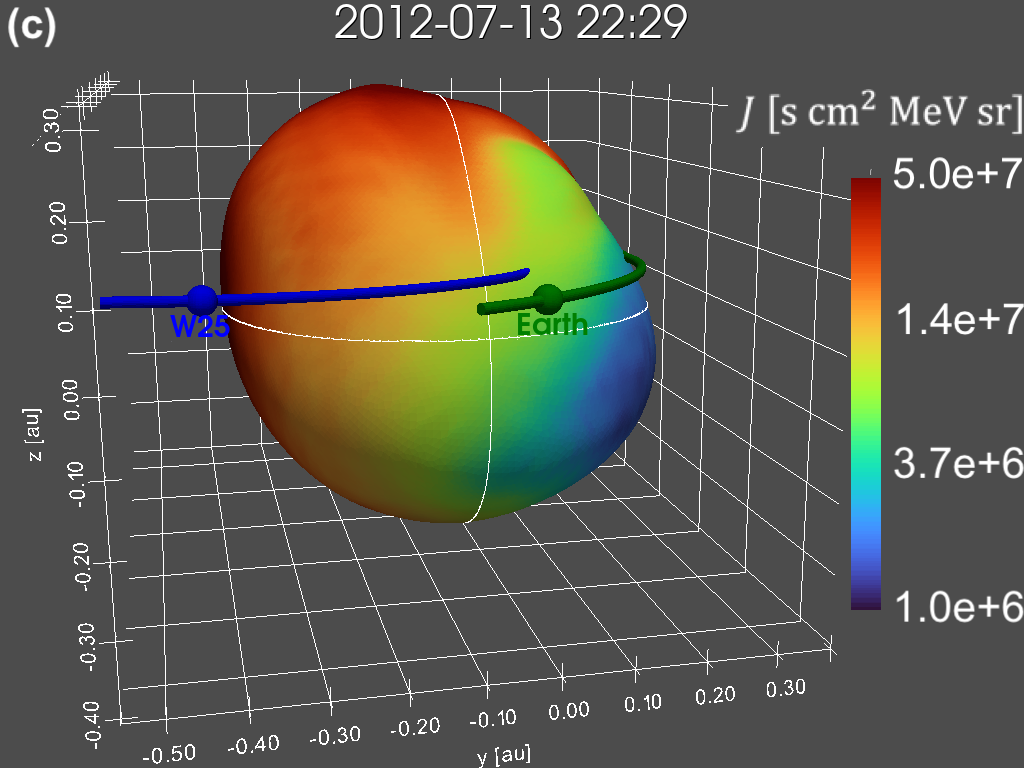}
    \hspace{-2.00mm}
    \includegraphics[width=0.49\textwidth]{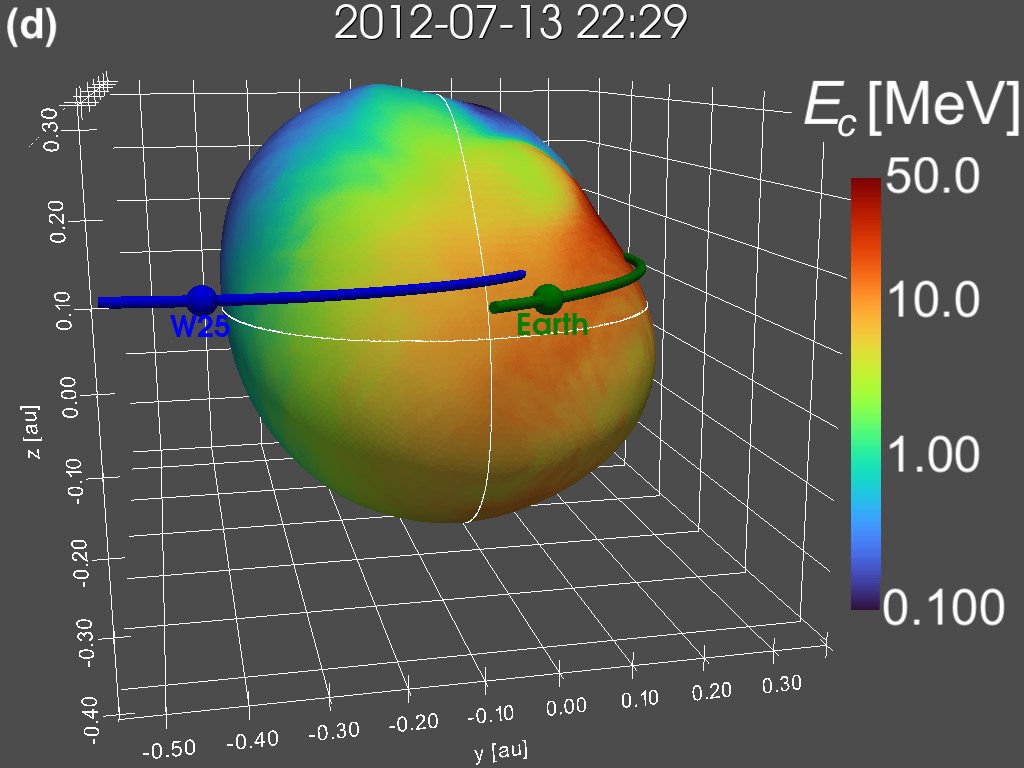}
    \caption{The CME-driven shock, observed 27 h after its introduction into the simulation domain, is depicted. The shock surface is color-coded to represent various parameters: the shock obliquity $\theta_{Bn}$ (in panel (a)); the Mach Alfvén number $M_\mathrm{A}$ (in panel (b));  the intensity $J$ and defined in Eq. \ref{eq:part spec at shock} (in panel (c)); and the cutoff energy $E_\mathrm{c}$, as defined in Eq. \ref{eq:cutoff energy fit} (in panel (d)). Green and blue dots indicate Earth and the W25 observer, respectively, and their magnetic connections to the shock wave are also illustrated. The white lines on the shock indicate the location of the solar equatorial plane and the meridional slice containing Earth.}
    \label{fig:sim_shock}
\end{figure}

These energy-dependent variations of the SEP intensities along the shock surface can be better understood by examining the properties of the shock, as illustrated in Figure \ref{fig:sim_shock}, which corresponds to the same date and time as the intensity snapshots in Figure \ref{fig:sim_slice}. Panels~a and~b of Figure \ref{fig:sim_shock} depict the shock angle $\theta_{B_n}$ and the Alfvenic Mach number $M_\mathrm{A}$, respectively. One key observation is that the eastern flank of the CME exhibits quasi-parallel shock geometry ($\theta_{B_n} < 45^\circ$), while the center and the western flank of the CME display quasi-perpendicular shock geometry ($\theta_{B_n} > 45^\circ$). This spatial variation in the shock geometry is a direct consequence of the spiral geometry of the upstream IMF. Additionally, the Alfvenic Mach number $M_\mathrm{A}$ reaches its maximum value at the center and the western flank of the CME shock. This is attributed partly to the shock being quasi-perpendicular in these regions (recall that $M_\mathrm{A}$ is measured in the HT-frame) and partly due to variations in the non-uniform upstream density and solar wind speed (not shown in the figure).

These spatial variations translate into different levels of particle emission and energy spectra in the PARASOL model at the shock wave. To illustrate this, we present the intensity parameter $J$ and the cut-off energy $E_c$ in panels c and d of Figure \ref{fig:sim_shock}. As indicated in Eq. \eqref{eq:part spec at shock}, $J$ represents the energy-independent part of the particle intensity at the shock wave. Figure \ref{fig:sim_shock}c clearly illustrates that the highest values for $J$ are concentrated in the east flank of the shock wave. However, the same eastern flank exhibits relatively lower values for the cut-off energy $E_\mathrm{c}$, compared to the nose and the west flank of the CME (see Figure \ref{fig:sim_shock}d).

In summary, the varying shock conditions along the shock surface explain why the eastern shock flank generates the highest intensities in the low-energy channels, while the west flank produces the highest intensities in the high-energy channels. The intricate interplay of shock geometry, Alfvenic Mach number, and upstream conditions plays thus a pivotal role in shaping the energy-dependent patterns in the distribution of  proton intensities.

\section{Discussion} \label{sec:discussion}
A test PARASOL simulation of the July 12, 2012 SEP event using the EUHFORIA MHD simulation of the solar wind and the CME have reproduced the observed ESP event ($E \lesssim 5$~MeV) in the close vicinity of the shock reasonably well (within one order of magnitude in intensity). The overestimated intensities at higher energies are due to an overestimated cutoff energy at 1~au in the simulation, which is governed by Eq.~\eqref{eq:theory cutoff energy} and largely results from the assumed $\Delta t = 10$~h. It should be noted that this parameter is not necessarily constant and can be a function of heliocentric distance. 

In the particle acceleration model where the shock upstream turbulence is self-generated, one of the possibilities is that the cutoff energy is ultimately governed by the flux of protons injected into the acceleration process. In this case, a decrease in the plasma density in the interplanetary medium should lead to a decrease in the injected proton flux (assuming that $\epsilon_\mathrm{inj}$ does not change much). This should lower the cutoff energy while the shock propagates further away from the Sun, if the maximum cutoff energy was achieved when the shock was in the corona. In this case, to constrain the parameter $\Delta t$ as a function of distance from the Sun one needs to use a SOLPACS simulation in a global setup, i.e., taking into consideration heliospheric variations of the solar wind plasma density and the magnetic field. 
Another option (perhaps more applicable for strong shocks) is that the cutoff energy is controlled by the particle escape from the foreshock due to magnetic focusing \citep{Vainio14}. In this case, it can be shown that $\Delta t$ does not enter the equations determining the cutoff energy. This will be addressed in a future study.

As we mentioned above, the ESP event under consideration was modeled earlier by \cite{wijsen22}, in which PARADISE was used not only to propagate but also to accelerate particles (50~keV seed protons) at the EUHFORIA-modeled CME shock. The study employed in-situ observations of the ESP event to estimate the spatial variation of the mean free path in the foreshock region. In contrast, the new PARASOL model does not rely on SEP observations for the event, making it more suitable for forecasting purposes.

The energy spectrum at the shock in \cite{wijsen22} showed a strong agreement with observations up to 1 MeV. However, above 1 MeV, the spectrum was too soft. The authors attributed this softening to low SEP acceleration efficiency at the modeled MHD shock as a result of the finite resolution of the MHD simulation leading to an overly thick shock. With PARASOL, instead of modeling particle acceleration with PARADISE, we inject an accelerated particle population based on MHD shock properties and the SOLPACS-based model. This approach circumvents the issue of shock thickness due to finite resolution, but makes the cutoff energy to depend on a free parameter ($\Delta t$) of the model. 

A striking discrepancy between the observed and simulated time-intensity profiles in Fig.~\ref{fig:obs_and_sim}, not related to the EUHFORIA simulation of the solar wind and CME, can be seen in the low energy channels (up to $\sim$0.5~MeV) before the shock arrival. The observed time-intensity profiles at these energies nearly overlap, which corresponds to a nearly flat energy spectrum of particles, starting from some distance from the shock in the upstream region. This feature is further illustrated in Fig.~\ref{fig:energy_spectrum_sim}b, which shows the energy spectra eight hours before the shock arrival, at the onset of the ESP event. Here, the observed energy spectrum is particularly hard (nearly constant) below 1 MeV. In contrast, in the simulations, intensities at low energies decrease rapidly with an increasing distance from the shock in the upstream region, resulting in an ``inverted" energy spectrum (in which intensity increases with energy) at a given location in the upstream region. The formation of such a spectrum in the PARASOL simulation stems from the SOLPACS self-consistent simulations. Namely, in SOLPACS, higher-energy protons are able to generate Alfv\'en waves in the upstream region at frequencies resonant with lower-energy protons, thus leading to a decrease of the scattering mean free path at the lower energies. This leads to an increasingly efficient trapping of lower-energy protons close to the shock, and may result (depending on how efficiently the waves are produced) in an ``inverted" energy spectrum of protons in the upstream region. 

Overlapping particle intensities upstream of ESP events is not uncommon, as recently pointed out by \citet{Lario18}.
However, the mechanism responsible for such flat, rather than inverted, energy spectrum at low-energy particle intensities is currently not understood, but some ideas have been proposed.  Those include a balance between the particle acceleration at the shock and adiabatic cooling \citep{Prinsloo19} and a velocity filter mechanism \citep{Perri23}, which suggests that the lower the particle energy, the smaller the fraction of such particles that can escape from the shock, assuming an isotropic distribution of particles.
Clearly, the inverted spectrum situation obtained in our simulations is a result of too efficient trapping of particles near the shock due to too small mean free paths of low-energy particles. If the mean free paths at low energies were larger, the intensities of particles at those energies would be larger (the ``inverted" part of the spectrum would be flatter). Importantly, the mean free path in SOLPACS (and PARASOL) is an increasing function of particle energy. 

Earlier, \cite{wijsen22} calculated the mean free paths in this ESP event from the observed particle intensities and found that they increase with energy only in a close vicinity of the shock ($\lesssim 1.7~R_\odot$ along the normal to the shock), where the intensities are not overlapped, but further away, where the intensities become overlapped, the energy dependence of the mean free path turns over. Moreover, the mean free paths at the lowest energies further away from the shock are substantially (about one order of magnitude) larger than close to the shock. 

The rigidity dependence of the mean free path in the quasi-linear theory is determined by the power-law index $q$ of the turbulence spectrum $I(k)$, $\lambda \propto P^{2-q}$ \citep[e.g.,][]{Bieber94}. In SOLPACS, the spectrum of self-generated Alfv\'en waves is characterized by power-law indices $q \lesssim 2$, which explains the resulting energy dependence of the mean free path. However, a steeper turbulence spectrum ($q>2$) would result in an opposite energy dependence of the mean free path (the mean free path decreases with particle energy). One candidate process, not accounted for in the SOLPACS model, that can produce such a steeper spectrum at the high end of the MHD frequency range is the Alfv\'en wave damping on thermal protons \citep[e.g.,][]{Berezhko16}. The spectrum steepening, however, produces a resonance gap \citep{Smith1990} and makes it more difficult for particles to scatter across the pitch angle of $90^{\circ}$, which should lead toward reduction of the particle acceleration efficiency at the shock. The effect of resonance broadening is usually invoked to deal with this problem, which allows a particle to interact not only with the wave spectral component strictly determined by the quasi-linear resonance condition \citep[e.g.,][]{Vainio03}, but with spectral components from a wider range of wavenumbers \citep[see, e.g.,][and references therein]{Bieber94, Ng95}. 

\cite{Ng95} proposed a theory where the resonance broadening was attributed to large-amplitude medium-scale fluctuations of the magnetic field, non-resonant with the particles at the considered energies. Bearing that theory in mind, we hypothesize that the resonance broadening can be very efficient close to the shock (stronger large-amplitude magnetic field variations), so that it allows large pitch-angle particles to interact with lower-frequency components of the self-generated Alfv\'en wave spectrum not affected significantly by the damping process. It can be shown that this can lead to a mean free path increasing with energy. Further away from the shock, the resonance broadening becomes less efficient, so the effect of a steeper spectrum due to the wave damping becomes more important, thus leading to the opposite dependence of the mean free path on energy. Also, one can expect in this case that the mean free path values, especially at lower energies, should be larger in magnitude further away from the shock than close to the shock. This scenario qualitatively agrees with the theory of \citet[][see their Figure 9]{Ng95}. 

It is also interesting to note that in some events where the presence of the flat spectrum was reported, the spectra are in fact slightly inverted than flat \citep[see, e.g., Fig.~4 in][]{Ng12}. Besides, the significantly inverted particle spectrum in a gradual SEP event has been recently reported by \cite{Lario21} and \cite{Ding24}. In the latter work, such a spectrum was attributed to the presence of a pre-existing enhanced turbulence in the ambient solar wind.   

It would be interesting to test the described scenario of the formation of overlapped intensities (nearly flat spectra) in the shock upstream region. This can be done by including the effect of Alfv\'en wave damping on thermal protons to the SOLPACS model and considering the resonance broadening as decreasing with the distance from the shock (towards upstream). A successful reproduction of this phenomenon in SOLPACS simulation would allow us to fine-tune our semi-analytical model of the inner foreshock in order to model (and eventually predict) this behavior of particle intensities at lower energies.     

\section{Conclusions and Outlook} \label{sec:Conclusions}
In this paper, we introduced a new physics-based simulation model, PARASOL, for simulating proton acceleration in a CME-driven shock and their transport in the interplanetary medium. PARASOL combines a semi-analytical model of the inner foreshock region, derived from self-consistent simulations of particle acceleration using the SOLPACS model, with the test-particle simulation model, PARADISE, which handles particle transport in the outer foreshock and the ambient solar wind, where wave growth is not significant. PARASOL requires a MHD model of the solar wind and the shock in order to obtain the necessary plasma and shock parameters. In particular, the semi-analytical model of the foreshock requires the ambient plasma density, $n$, the Alfv\'en speed, $V_\mathrm{A}$, the shock-normal angle, $\theta_\mathrm{Bn}$, all at the shock location, and the Alfv\'enic Mach number (in the HT frame) $M_\mathrm{A}$ of the shock. For a test PARASOL simulation, we used a EUHFORIA simulation of the CME associated with the July 12, 2012 SEP event. 

To constrain better the free parameters of PARASOL, the focus of a future modeling study should be on ESP events having the so-called classic component, i.e., a slow increase of the particle intensity starting a few hours before the shock arrival \citep{Lario2003, Giacalone2012}. Such a rise of intensity is consistent with the presence of the foreshock.

The ongoing development of PARASOL aims to create an efficient simulation model for operational SEP forecasting that also provides an accurate description of the physical processes involved in particle acceleration, particularly Alfvén wave growth upstream of the shock. The first version of the model presented here focuses mainly on the ESP component of an SEP event. Currently, the entire simulation chain takes around 1500 CPU hours. On a supercomputer, this allows us to produce a forecast within roughly 2 hours, which is adequate for predicting an ESP event in a timely manner. However, there are several potential optimizations that could significantly reduce both the computational and prediction times. For instance, EUHFORIA and PARASOL are currently run sequentially, but they could be executed in parallel to save time. Additionally, in the simulations presented here, particles were injected along the entire shock wave, which adds considerable computational cost. In a near-Earth forecasting scenario, it would be sufficient to inject particles only in the vicinity of a shock location that is magnetically connected to Earth, thus reducing the necessary CPU hours drastically.

To address a full SEP event, such as capturing the event onset, an MHD model of the propagating shock in the corona is required. We are working on such a simulation framework by combining EUHFORIA with the recently developed coronal model COolfluid COroNal UnsTructured (COCONUT) \citep[][]{Perri23}, which can perform self-consistent flux-rope simulations \citep[see e.g.,][]{linan2023}. This integrated approach promises to enhance the accuracy and reliability of SEP event predictions significantly.

\begin{acknowledgements}
This research has received funding from the European Union’s Horizon 2020 research and innovation programme under grant agreements No 870405 (EUHFORIA 2.0). 
A.A.\ and R.V.\ also acknowledge funding from the European Union’s Horizon 2020 research and innovation programme under grant agreement No 101004159 (SERPENTINE) and from the Finnish Centre of Excellence in Research of Sustainable Space (FORESAIL) funded by the Research Council of Finland (grant no.\ 352847). The study has been performed in the framework of the European Union’s Horizon 2020 research and innovation programme (EU/H2020) under the Marie Skłodowska-Curie grant agreement No.\ 955620 (SWATNet). 
N.W.\ acknowledges funding from the Research Foundation -- Flanders (FWO -- Vlaanderen, fellowship no.\ 1184319N) and the KU Leuven project 3E241013. Computational resources and services used in this work were provided by the Finnish IT Center for Science (CSC) and the FGCI project (Finland) and by the VSC (Flemish Supercomputer Centre), funded by the FWO and the Flemish Government-Department EWI. 
\end{acknowledgements}

%%    This version assumes use of bibtex with the jswsc.bib file being present
%%    If your bib file has a different name you need to change the following line

\bibliography{references}

\Online

\begin{appendix} 
\section{Scaling of SOLPACS equations} \label{appendix scaling property}

In Eq.~\eqref{eq:solpacs_particle_eq}, the particle distribution function 
%at the injection momentum $f(p_\mathrm{inj})$ 
$f$ scales as $\epsilon_\mathrm{inj} n$, where $\epsilon_\mathrm{inj}$ is the model parameter describing the efficiency of the shock to inject low-energy particles to the acceleration process \citep[for details, see, e.g.,][]{Afanasiev23}. Taking this into account and using a scaled wavenumber $k^{*}=k/B$ in Eq.~\eqref{eq:wave_grw}, we infer that 
%at the moment of initial injection 
$\Gamma\left(k\right)=\Gamma\left(k^{*}\right)\propto\epsilon_{\mathrm{inj}}\sqrt{n}$. We note also that $\Gamma$ does not depend explicitly on $B$.

Furthermore, one can write for a power-law wave spectrum (or for a portion of the wave spectrum approximated by a power law):
\begin{equation}
    I\left(k\right) = I_{0}\left(\frac{k}{k_{0}}\right)^{-q_0} = I_{0}\left(\frac{k^{*}}{k_{0}^{*}}\right)^{-q_0} \equiv I^{*}\left(k^{*}\right).
\end{equation}
%where the normalization factor $I_0$ is obtained using the definition of the particle mean free path
%\begin{equation}
%    \lambda = \frac{3v}{8}\int_{-1}^{1}\frac{(1-\mu^2)^2}{D_{\mu\mu}}d\mu.
%\end{equation}
%as
%\begin{equation}
%    I_0 = \frac{3}{\pi(2-q_0)(4-q_0)}\left(\frac{\gamma v_0}{\Omega_{0}^{*}}\right)^2\frac{1}{\lambda_0},
%\end{equation}
%where $\Omega_0^{*} = e/m_\mathrm{p} = \Omega_0/B$, and $\lambda_0$ is the mean free path of protons having speed $v_0$. 
Therefore, using the scaled wavenumber $k^{*}$, the pitch-angle diffusion coefficient can be given as 
\begin{equation}
    D_{\mu\mu}=\frac{\pi}{2}\Omega_{0}^{*}\frac{\left|k_{\mathrm{r}}^{*}\right|I^{*}\left(k_{\mathrm{r}}^{*}\right)}{\gamma}\left(1-\mu^{2}\right),
\end{equation}
where $\Omega_0^{*} = \Omega_0/B = e/m_\mathrm{p}$, so it does not depend on the magnetic field magnitude. 
Thus, under any transformation of the plasma and shock parameters such that the parameters $M_{\mathrm{A}}$, $V_{\mathrm{A}}$ and $\epsilon_{\mathrm{inj}}\sqrt{n}$ are constant, Eqs.~\eqref{eq:solpacs_particle_eq} and \eqref{eq:solpacs_wave_eq} are invariant. We also have to require the constancy of $\theta_\mathrm{Bn}$ in order to have invariant boundary conditions for particles at the shock. This transformation can also be summarised as follows:  
\begin{flalign*}
    M_{\mathrm{A}} & = \mathrm{const}.\\
    V_{\mathrm{A}} & = \mathrm{const}.\\
    \theta_\mathrm{Bn} & = \mathrm{const}.\\
    n & \rightarrow \alpha n\\
    \epsilon_{\mathrm{inj}} & \rightarrow \alpha^{-1/2}\epsilon_{\mathrm{inj}},
\end{flalign*}
where $\alpha$ is a constant. Note that while the particle equation is invariant, the distribution function of accelerated particle being proportional to $\epsilon_{\mathrm{inj}}n$ scales under this transformation as $\alpha^{1/2}$. 

\begin{figure}
    \centering
    \includegraphics[scale=0.4]{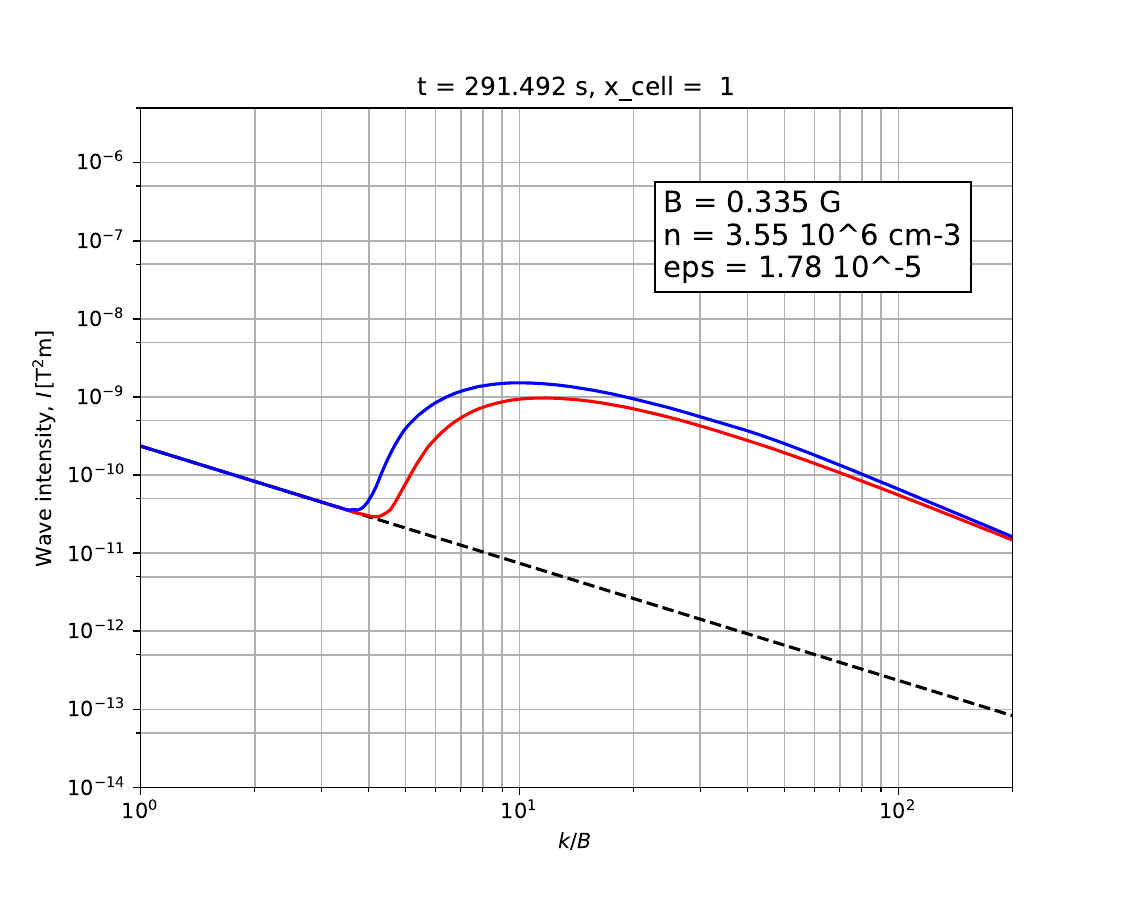}
    \includegraphics[scale=0.4]{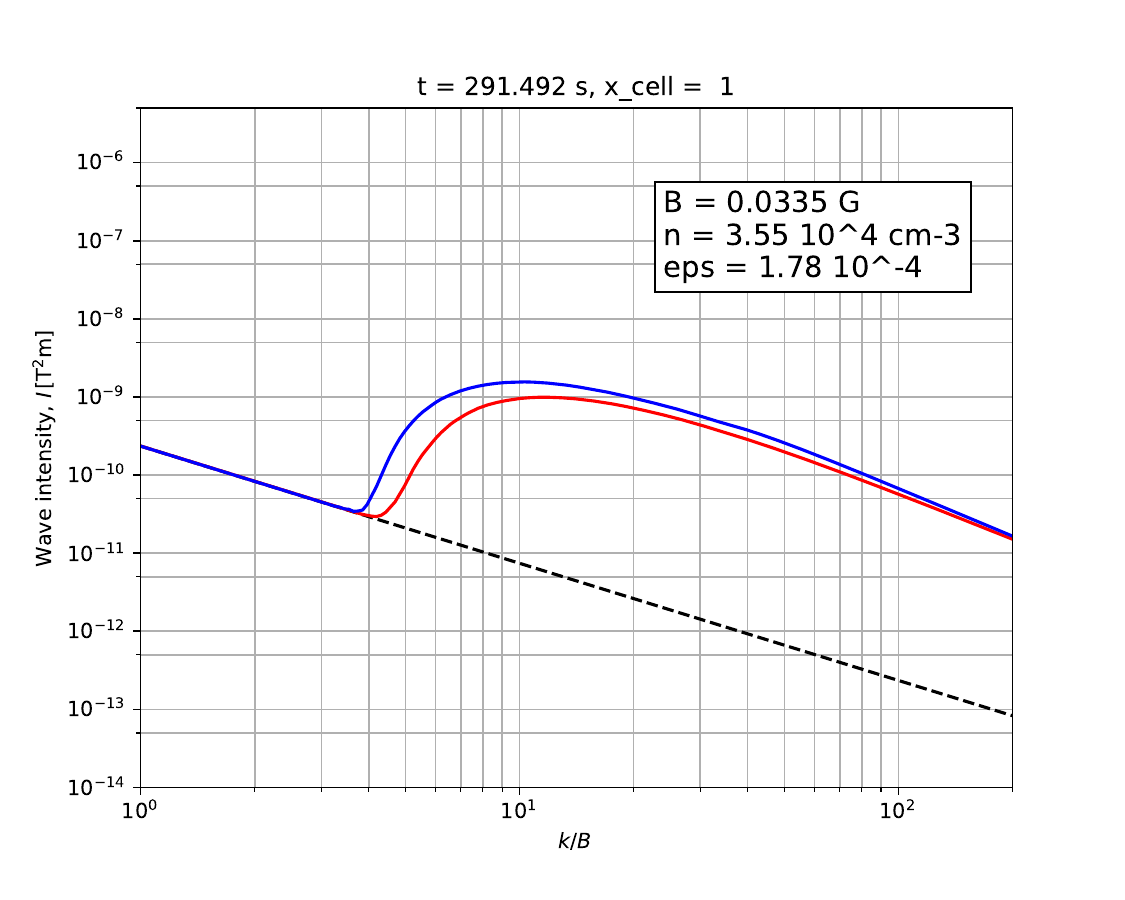} \\
    \includegraphics[scale=0.4]{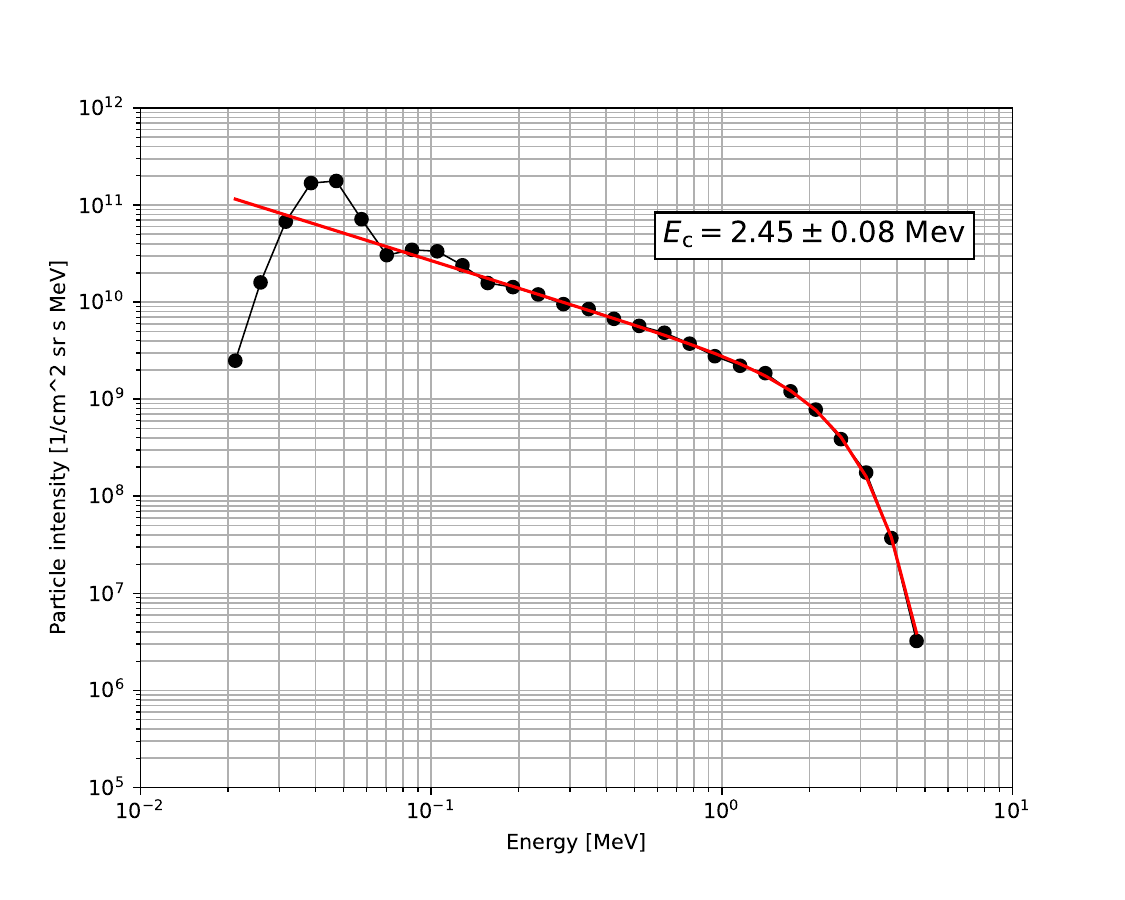}
    \includegraphics[scale=0.4]{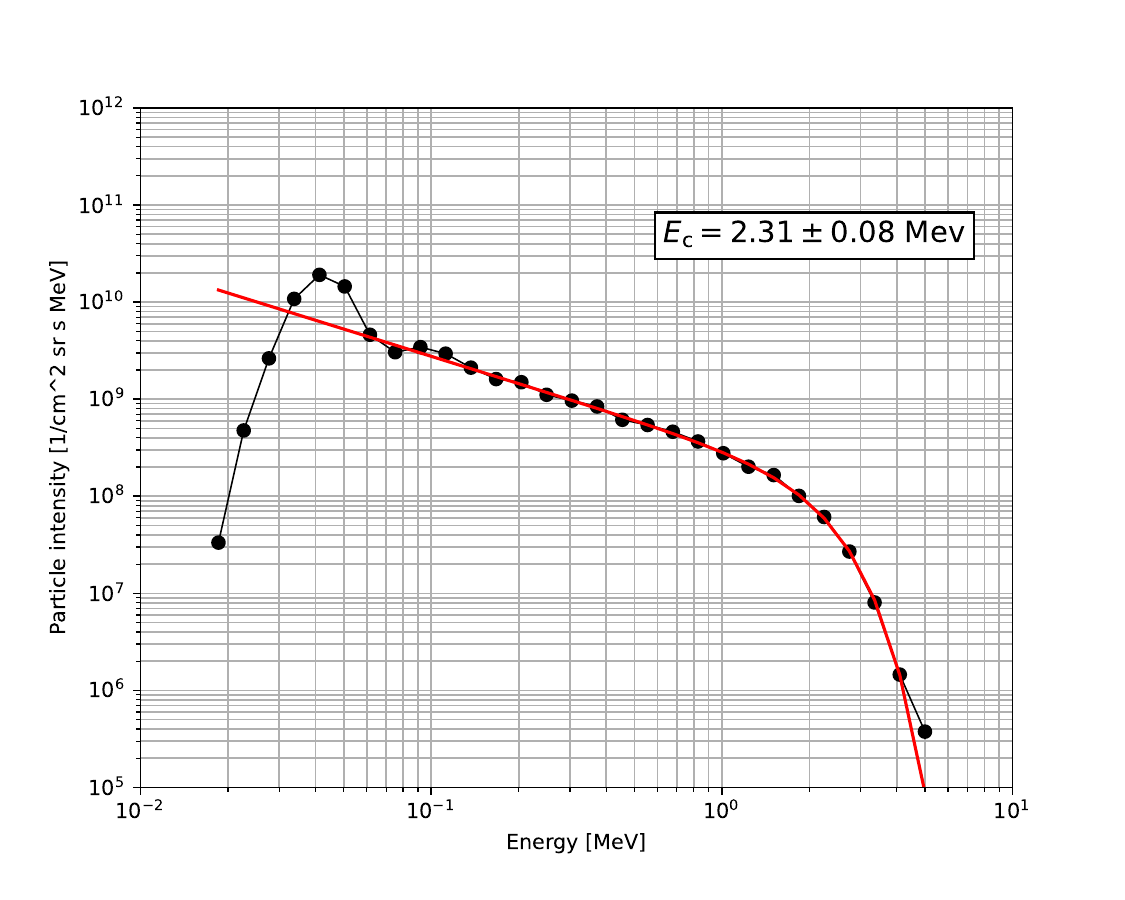}
    \protect\caption{Comparison of the Alfv\'{e}n wave spectra (the blue and red lines depict the opposite circular polarizations) in the vicinity of the shock (top row) and the proton spectra at the shock (bottom row) resulting from two simulation runs to test the scaling property of the SOLPACS equations.
    \label{fig:scaling test}}
\end{figure}

We have performed test simulations to confirm the discovered scaling property. Figure~\ref{fig:scaling test} shows an example of testing this transformation property for $B = 0.335$~G, $n = 3.55 \cdot 10^6$~cm$^{-3}$, $V_\mathrm{sh} = 1500$~km/s, $\theta_\mathrm{Bn} = 0$, and $\alpha = 0.01$. It can be seen that, as expected, the wave spectrum as a function $k/B$ is invariant, and the particle intensity scales as $\alpha^{1/2} = 0.1$ while the spectral shape is preserved.  

\end{appendix}

\begin{appendix}
\section{Bell's theory} \label{appendix: Bell's theory}

The 1D steady-state theory of particle acceleration in a parallel shock accounting for self-generated Alfv\'{e}n waves \citep{Bell78} can be formulated as follows \citep[see also][]{Vainio14, AfanasievVainio15}: 
\begin{equation}
f(x,p)=f_{\mathrm{s}}\frac{x_{0}}{x+x_{0}}, \label{eq:bell part func}
\end{equation}
\begin{equation}
\lambda(x, v) = \frac{C_\lambda}{v} (x + x_0), \label{eq:bell mfp}
\end{equation}
\begin{equation}
f_{\mathrm{s}}(p) = C_f \left(\frac{p}{p_{\mathrm{inj}}}\right)^{-\sigma},\label{eq:bell part func shock}
\end{equation}
\begin{equation}
x_{0}(p) = C_x \left(\frac{p}{p_{\mathrm{inj}}}\right)^{\sigma-3}, \label{eq:bell x0}
\end{equation}
where $f$ is the particle distribution
function, $\lambda$ is the scattering mean free path, $v$ and $p$ are the particle speed and momentum, 
$\sigma=3r_{\mathrm{c}}/(r_{\mathrm{c}}-1)$
is the spectral index of the particle distribution function at shock,
$r_{\mathrm{c}} = r_\mathrm{g} \left(1 - M^{-1}_\mathrm{A}\right)$ is the scattering-centre compression
ratio, $r_\mathrm{g}$ is the gas compression ratio of the shock, 
$p_{\mathrm{inj}}$ is the effective injection momentum \citep[see][for details]{AfanasievVainio15}, and
\[
C_\lambda = 3 \left(u_1 - V_\mathrm{A}\right), 
\quad C_f = \frac{\sigma\epsilon_{\mathrm{inj}}n}{4\pi p_{\mathrm{inj}}^{3}}, 
\quad C_x = \frac{2}{\pi\epsilon_{\mathrm{inj}}\sigma}\frac{V_{\mathrm{A}}}{\Omega_{\mathrm{0}}},
\]
are the functions of the plasma and shock parameters, which are constant in the context of local SOLPACS simulations.
\end{appendix}
   
%\end{linenumbers}

\end{document}